\documentclass[fleqn,usenatbib]{mnras}

\usepackage{newtxtext,newtxmath}
\usepackage[T1]{fontenc}
\usepackage[normalem]{ulem}

\DeclareRobustCommand{\VAN}[3]{#2}
\let\VANthebibliography\thebibliography
\def\thebibliography{\DeclareRobustCommand{\VAN}[3]{##3}\VANthebibliography}

\usepackage{graphicx}	% Including figure files
\usepackage{amsmath}	% Advanced maths commands
\usepackage{pdflscape}
\usepackage{float}

\title[Evolution of young stellar clusters in Orion]{Dynamical evolution and dissolution timescale of young stellar clusters in the Orion star-forming complex}

\author[S. Sánchez-Sanjuán et al.]{
Sergio Sánchez-Sanjuán,$^{1}$\thanks{E-mail: sergsanchez@astro.unam.mx}
Ángeles Pérez-Villegas,$^{1}$
Jesús Hernández$^{1}$ and
Luis Aguilar$^{1}$
\\
$^{1}$Universidad Nacional Autónoma de México. Instituto de Astronomía. A.P. 106, 22800. Ensenada, B.C., México
}

\date{Accepted XXX. Received YYY; in original form ZZZ}

\pubyear{\the\year{}}

\begin{document}
\label{firstpage}
\pagerange{\pageref{firstpage}--\pageref{lastpage}}
\maketitle

% Abstract of the paper
\begin{abstract}

We present a comprehensive analysis of the Orion star-forming complex (OSFC), combining structural, kinematic, and dynamical information to constrain the present-day state and future evolution of its stellar substructures. Using \textit{Gaia} DR3 astrometry and complementary radial velocities from high-resolution spectroscopic surveys, we derived three-dimensional velocity distributions and structural parameters for 13 young clusters. For the stellar component, we estimated a correction of the present-day mass function for observational incompleteness and calculated the virial state, $\alpha_{\rm vir}$, finding that all clusters are supervirial. Direct $N$-body simulations initialized from the present-day global parameters and evolved for 300~Myr in a Galactic potential suggest a separation of the OSFC clusters into two regimes: seven clusters with $\alpha_{\rm vir}\lesssim 7$ evolve in a Galactic-potential-regulated regime that retains a bound core for $\gtrsim 170$ Myr as long-lived open clusters, whereas six clusters with $\alpha_{\rm vir}\gtrsim 7$ enter an internal-dynamics--dominated regime, dissolving before 120 Myr and likely populating the Galactic stellar field. For both regimes, a control test indicates negligible cluster--cluster interactions under current OSFC conditions. Finally, long-lived clusters show low-amplitude modulations in the bound fraction correlated with the Galactic vertical motion, consistent with disk-crossing tidal heating and the temporary recapture of marginal members. These results highlight the OSFC as a natural laboratory where heterogeneous initial conditions give rise to persistent open clusters and dispersing groups.

\end{abstract}

% Select between one and six entries from the list of approved keywords.
% Don't make up new ones.
\begin{keywords}
stars: pre-main-sequence -- stars: kinematics and dynamics -- open clusters and associations: individual: Orion OB1, $\lambda$~Orionis -- methods: statistical

\end{keywords}

%%%%%%%%%%%%%%%%%%%%%%%%%%%%%%%%%%%%%%%%%%%%%%%%%%
%%%%%%%%%%%%%%%%% BODY OF PAPER %%%%%%%%%%%%%%%%%%

\section{Introduction}

%%%
Star-forming complexes are the largest coherent sites of recent star formation in the Milky Way, arising from the fragmentation and gravitational collapse of giant molecular clouds~\citep{kroupa:2001, elmegreen:2002, lada:2003, krumholz:2019}. They contain a hierarchy of stellar groupings, ranging from compact to extended clusters, which preserve the kinematic imprint of their natal environments during the early stages of stellar evolution~\citep[e.g.][]{vazquez-semadeni:2017, wright_mamajek:2018, krause:2020}. Their study provides insight into the star formation process, the structure of the Galactic disc, and the Milky Way star formation history~\citep{zari:2018, beccari:2018, dalessandro:2021, kerr:2023}.

With the advent of \textit{Gaia}~\citep{gaia:2016, gaia:2018, gaia:2021}, observational studies have revealed that most stellar systems often exhibit dynamical signatures of expansion and dissolution~\citep{kounkel_covey:2019, kuhn:2019, cantat-gaudin:2019, armstrong:2020, quintana:2023}. Such expanding stellar systems correspond to OB associations~\citep{ambartsumian:1947, blaauw:1964}, characterised by low stellar densities and dispersal over relatively short timescales. Their gradual unbinding provides a physical link between the dense environments where stars are born and the diffuse stellar population of the Galactic disc~\citep{wright:2020}. Coherent velocity gradients, substructured phase-space distributions, and tidal features confirm that these systems are dynamically evolving rather than virialised~\citep{wright:2024}. Since clusters are not dynamically relaxed, structural parameters vary systematically with age~\citep{portegies:2010}.

In particular, the Orion star-forming complex (OSFC), located at a distance of $\sim$400 pc~\citep{grossschedl:2018, zucker:2020}, is one of the closest and most massive star-forming regions, making it an ideal laboratory for studying cluster dynamics across a range of evolutionary stages between 1 and 20 Myr~\citep{bally:2008, kounkel:2018, koss:2019}. The OSFC exhibits complex spatial and kinematic substructure, with evidence of a global expansion in Orion OB1, ballistic expansion in $\lambda$ Orionis, and signatures of past interactions between subclusters~\citep{kounkel:2018, zari:2019, swiggum:2021, pang:2021, sansan:2024}. Despite this wealth of kinematic information, the long-term dynamical evolution of the Orion complex remains poorly constrained, as few studies have attempted to link observational characterisations directly with predictive dynamical models.

The early evolution of young stellar clusters (YSCs) is strongly shaped by residual gas expulsion, driven by winds, ionising radiation, and supernovae from massive stars~\citep{hills:1980, baumgardt:2007, goodwin:2006, kruijssen:2012, zamira-aviles:2019, krause:2020}. This phase, occurring within the first $\sim$10 Myr, often leads to significant expansion or even dissolution. Observations suggest that up to 80\% of clusters younger than 30 Myr show signs of ongoing expansion~\citep{dellacroce:2024}. Those that remain bound subsequently evolve under the influence of internal processes such as stellar evolution, mass loss, two-body relaxation, and mass segregation, as well as external perturbations from the Galactic tidal field and encounters with giant molecular clouds~\citep{spitzer:1987, gieles:2006, lamers:2005}.

Numerical simulations have shown that cluster lifetimes depend critically on their initial mass, density profile, and Galactocentric distance~\citep{baumgardt:2007, portegies:2010, ernst:2011, wang:2016, sills:2018}, with early expansion producing dissolution timescales ranging from a few tens of Myr for low-mass, diffuse groups to several Gyr for massive, compact clusters~\citep{baumgardtMakino:2003, dinnbier:2022, farias:2023}. Therefore, a precise characterisation of cluster properties is needed to anchor realistic dynamical predictions. %for star-forming complexes.

Previous studies have begun linking observations with the dynamical evolution of YSCs, but with important limitations. \citet{wright:2024} characterized the spatial and kinematic structure of 18 nearby clusters without extending the analysis to predictive dynamical modelling. \citet{pang:2021_2} combined \textit{Gaia} data with $N$-body simulations to study the expansion of Vela OB2, but neglected the external Galactic potential and did not quantify the dissolution times. Cluster-specific efforts, such as the \textit{Gaia}-ESO modelling of $\gamma$ Velorum \citep{mapelli:2015}, have successfully reproduced observed kinematic substructures, but were restricted to single systems. More recently, \citet{grossschedl:2025} examined the evolution of OB associations such as Sco–Cen from formation to the present day, without extending the dynamical modelling to predict the future dissolution.% of the observed clusters.

In the case of the OSFC, most simulation efforts have focused on reproducing the formation and early evolution of individual subregions, such as the Orion Nebula Cluster (ONC)~\citep[e.g.,][]{krumholz:2011, krumholz:2012, allison:2011, fujii:2021}. More recently, \citet{sills:2023} modelled the ONC, including a dominant gas component during its early embedded phase and assessing star formation up to 1 Myr. These studies generally adopt controlled or formation-motivated setups and do not follow their subsequent dynamical evolution after gas dispersal. To date, no work has combined the full set of \textit{Gaia}-derived properties of the OSFC with ensemble $N$-body simulations that include stellar evolution and the Galactic field to predict the future dispersal and survivability of the entire complex.

In this work, we investigate the dynamical evolution of stellar clusters in the OSFC using information from \textit{Gaia}-DR3 and high-resolution spectroscopic surveys. We derive structural and dynamical parameters using Bayesian inference, estimate the present-day virial states, and correct for observational incompleteness to recover the likely total stellar masses. Throughout this study, we consider only the stellar component, assuming a post-gas dynamical evolution scenario. These quantities are then used as initial conditions for an ensemble of $N$-body simulations that include stellar evolution and are evolved in a Galactic potential for 300 Myr. The paper is structured as follows. Section~\ref{sec:data} describes the data used in this study. Section~\ref{sec:clu_charac} details the methodology applied for the characterization of clusters. In Section~\ref{sec:pdmf}, we analyse the completeness of the stellar samples using the present-day mass function. Section~\ref{sec:sims} outlines the determination of initial conditions and the setup of the simulation ensemble. Section~\ref{sec:results} presents the main results and their statistical interpretation. Finally, Section~\ref{sec:conclusions} summarizes our conclusions.

\section{Observational Data}
\label{sec:data}

\begin{figure*}
    \centering
    \includegraphics[width=0.78\textwidth]{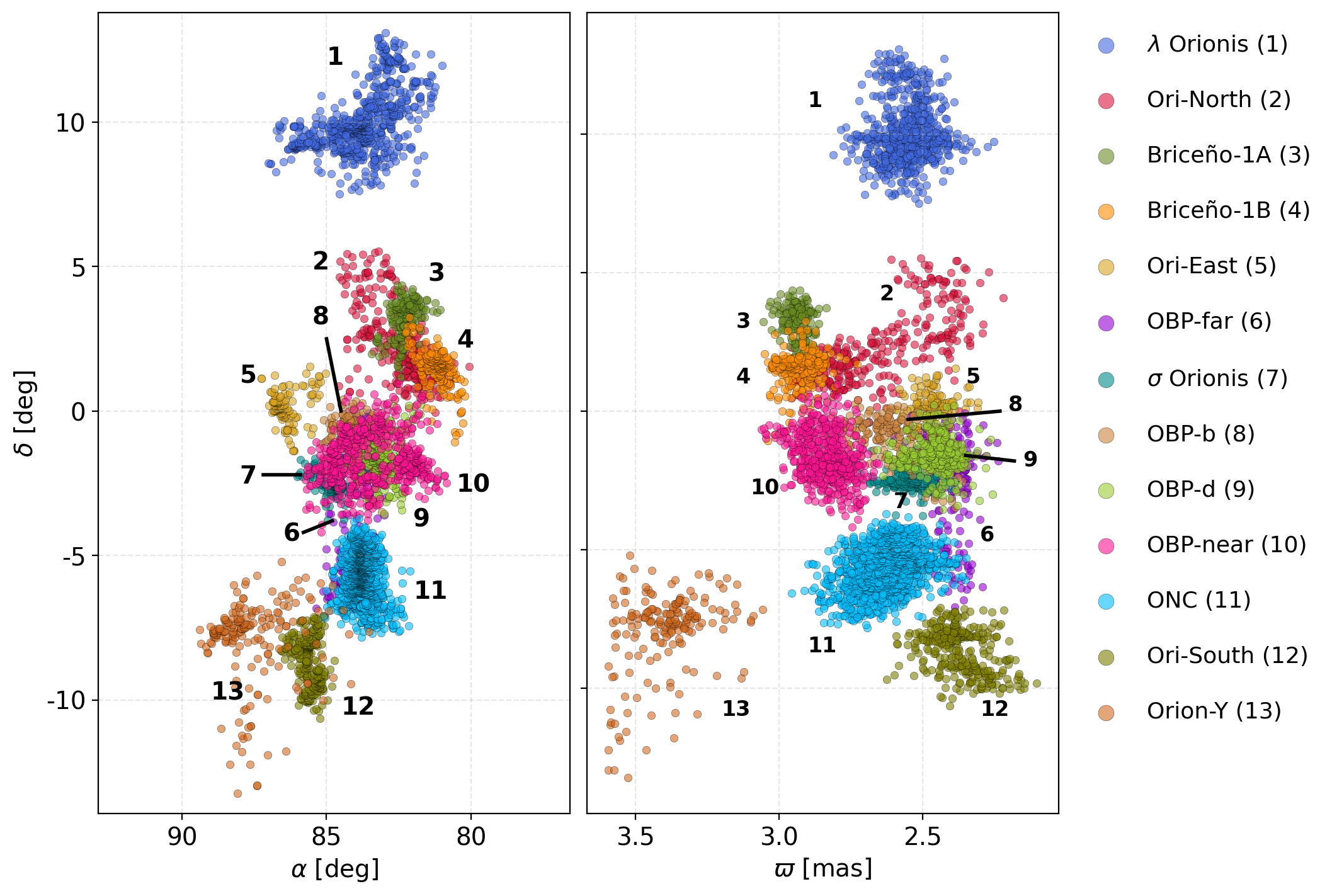}
    \caption{Spatial and kinematic distribution of confirmed stellar members in the OSFC. \textit{Left panel}: Equatorial coordinates ($\alpha$, $\delta$) showing the sky-plane morphology of each group. \textit{Right panel}: Parallax ($\varpi$) versus Declination. Labelled structures (1–13) are related to the identifier shown in Table~\ref{table:BS}.}
    \label{fig:clusters_BS}
\end{figure*}

For this work, we used the catalogue compiled by \citet{sansan:2024}, which identifies pre-main sequence (PMS) kinematic candidates with ages below 30 Myr in the OSFC. The clusters were found using a hierarchical clustering algorithm \citep[\texttt{HDBSCAN};][]{campello:2015}, applied to the 5D parameter space from \textit{Gaia} DR3. The initial sample contained 16,814 sources. The recovered clusters were classified into two regimes according to membership size: \textit{Big Structures} (more than 140 stars) and \textit{Small Structures} (fewer than 140 stars). The surveyed region spans $75^{\circ} < \alpha < 90^{\circ}$ and $-14^{\circ} < \delta < 16^{\circ}$, yielding 27 individual clusters.

All stars in the catalogue satisfy astrometric-quality criteria, based on parallax errors ($\sigma_\varpi/\varpi$), the renormalised unit weight error \citep[RUWE;][]{lindegren:2021}, and the fidelity parameter \citep{rybizki:2022}. Specifically, stars were required to meet $\sigma_\varpi/\varpi < 0.05$, and either $RUWE < 1.4$ or fidelity $> 0.5$. Additionally, for a sub-sample of star, the catalogue also provides radial velocities (RVs) from high-resolution spectroscopic surveys such as APOGEE-2 \citep{blanton:2017} and GALAH-DR3 \citep{buder:2021}, corrected by a systematic offset of $\sim$0.2~$\mathrm{km\ s^{-1}}$ between them. The fraction of stars with RV is provided in the last columns of Table~\ref{table:BS}.

For a dynamical analysis, we focused on clusters in the \textit{Big Structure} sample with at least 15\% of members having RV measurements, with one exception: Orion–Y, which has RVs for only 13 of 189 stars. Orion–Y also has a median parallax of 3.38 mas, significantly larger than the $\sim$2.5 mas typical of the other Orion groups, placing it well in the foreground of the complex. We therefore exclude Orion–Y and adopt as our working sample all remaining \textit{Big Structures}. Table~\ref{table:BS} summarises their median observational parameters, and Figure~\ref{fig:clusters_BS} shows the distribution of confirmed members.

{\renewcommand{\arraystretch}{1.24}
\begin{table*}
\centering
\caption{Mean astrometric and kinematic parameters of the OSFC \textit{Big Structures} derived from \textit{Gaia}-DR3. Reported uncertainties correspond to the 16th–84th percentile ranges of the distributions.}
\label{table:BS}
\begin{tabular}{clccccccc}
\hline
Cluster & \multicolumn{1}{c}{Name} & \begin{tabular}[c]{@{}c@{}}$\bar{\alpha}$\\ (deg)\end{tabular} & \begin{tabular}[c]{@{}c@{}}$\bar{\delta}$\\ (deg)\end{tabular} & \begin{tabular}[c]{@{}c@{}}$\bar{\varpi}$\\ (mas)\end{tabular} & \begin{tabular}[c]{@{}c@{}}$\bar{\mu}_\alpha^*$\\ ($\rm{mas}\ yr^{-1}$)\end{tabular} & \begin{tabular}[c]{@{}c@{}}$\bar{\mu}_\delta$\\ ($\rm{mas}\ yr^{-1}$)\end{tabular} & \begin{tabular}[c]{@{}c@{}}$\overline{RV}$\\ ($\rm{km}\ s^{-1}$)\end{tabular} & \begin{tabular}[c]{@{}c@{}}$N_T$ (with RV)\\ \end{tabular} \\ \hline \hline
1       & $\lambda$ Ori    & 83.75$^{+1.19}_{-1.01}$   & 9.83$^{+1.26}_{-0.72}$     & 2.55$^{+0.08}_{-0.07}$     & 1.25$^{+0.59}_{-0.59}$        & $-$2.12$^{+0.40}_{-0.33}$    & 27.26$^{+2.64}_{-1.63}$     & 773 (250)   \\
2       & Ori-North            & 82.39$^{+1.21}_{-0.71}$   & 2.09$^{+1.89}_{-1.05}$     & 2.62$^{+0.17}_{-0.24}$     & $-$0.67$^{+0.33}_{-0.34}$   & 0.73$^{+0.27}_{-0.29}$      & 30.63$^{+2.24}_{-1.70}$     & 305 (59)     \\
3       & Briceño-1A         & 82.34$^{+0.28}_{-0.41}$   & 3.43$^{+0.36}_{-0.92}$     & 2.94$^{+0.03}_{-0.04}$     & 1.53$^{+0.18}_{-0.20}$        & $-$0.59$^{+0.17}_{-0.15}$   & 20.35$^{+2.07}_{-2.67}$   & 175 (36)    \\
4       & Briceño-1B         & 81.13$^{+0.44}_{-0.34}$   & 1.62$^{+0.29}_{-0.61}$     & 2.93$^{+0.06}_{-0.05}$     & 1.38$^{+0.23}_{-0.17}$        & $-$0.03$^{+0.15}_{-0.19}$   & 20.83$^{+1.22}_{-1.40}$   & 237 (55)    \\
5       & Ori-East              & 86.54$^{+0.25}_{-0.50}$   & 0.13$^{+0.49}_{-0.44}$     & 2.47$^{+0.05}_{-0.05}$     & $-$0.72$^{+0.94}_{-0.58}$   & $-$0.83$^{+0.53}_{-0.44}$   & 28.11$^{+1.93}_{-1.42}$   & 144 (65)    \\
6       & OBP-Far             & 83.96$^{+0.56}_{-0.45}$   & $-$1.85$^{+0.98}_{-2.89}$   & 2.41$^{+0.06}_{-0.06}$    & $-$1.40$^{+0.51}_{-0.41}$   & 1.03$^{+0.36}_{-0.47}$      & 30.52$^{+3.05}_{-1.82}$   & 223 (31)    \\
7       & $\sigma$ Ori       & 84.78$^{+0.31}_{-0.23}$   & $-$2.55$^{+0.37}_{-0.19}$   & 2.54$^{+0.06}_{-0.06}$    & 1.46$^{+0.64}_{-0.39}$        & $-$0.47$^{+0.53}_{-0.43}$   & 30.84$^{+1.45}_{-1.98}$   & 207 (110)   \\
8       & OBP-b                & 84.00$^{+0.45}_{-0.54}$    & $-$0.85$^{+0.53}_{-0.90}$   & 2.61$^{+0.11}_{-0.14}$    & $-$1.02$^{+0.35}_{-0.27}$    & $-$0.64$^{+0.27}_{-0.20}$   & 31.12$^{+4.04}_{-5.40}$  & 196 (31)    \\
9       & OBP-d                & 83.15$^{+0.32}_{-0.41}$    & $-$1.65$^{+0.63}_{-0.40}$   & 2.45$^{+0.07}_{-0.05}$    & 0.08$^{+0.39}_{-0.24}$      & $-$0.21$^{+0.31}_{-0.26}$   & 30.84$^{+1.04}_{-1.00}$   & 301 (58)   \\
10      & OBP-Near         & 83.55$^{+1.20}_{-1.53}$    & $-$1.77$^{+1.07}_{-0.69}$   & 2.83$^{+0.08}_{-0.07}$    & 1.61$^{+0.28}_{-0.43}$      & $-$1.24$^{+0.24}_{-0.23}$   & 22.77$^{+2.06}_{-2.57}$   & 534 (133)    \\
11      & ONC                  & 83.82$^{+0.25}_{-0.30}$    & $-$5.58$^{+0.75}_{-0.84}$   & 2.62$^{+0.08}_{-0.08}$    & 1.16$^{+0.38}_{-0.44}$      & 0.28$^{+0.49}_{-0.74}$     & 26.98$^{+3.53}_{-2.55}$   & 1816 (828)     \\
12      & Ori-South           & 85.62$^{+0.27}_{-0.33}$    & $-$8.40$^{+0.48}_{-1.19}$   & 2.38$^{+0.07}_{-0.11}$    & 0.33$^{+0.38}_{-0.37}$      & $-$0.57$^{+0.54}_{-0.70}$   & 21.44$^{+1.80}_{-1.95}$   & 331 (117)    \\
13      & Orion Y              & 87.84$^{+0.70}_{-1.44}$    & $-$7.59$^{+0.70}_{-1.82}$    & 3.38$^{+0.14}_{-0.09}$    & 0.14$^{+0.32}_{-0.32}$     & $-$0.57$^{+0.24}_{-0.24}$    & 16.38$^{+2.42}_{-0.11}$   & 189 (13)     \\ \hline
\end{tabular}
\end{table*}
}

\section{Cluster Characterization}
\label{sec:clu_charac}

The dataset of likely OSFC members provides the observables required to derive present-day phase-space, structural parameters, masses, and virial states, which form the basis for dynamical modelling.
Since almost all clusters analysed here are observed as gas-poor stellar systems~\citep[e.g.][]{briceno:2005, kounkel:2018}, we consider only the stellar component as a suitable approximation~\citep[e.g.][]{mapelli:2015, pang:2021_2}. Once the natal gas has been largely dispersed, the long-term evolution is governed primarily by stellar self-gravity \citep{portegies:2010}. In clusters where some gas may remain (e.g. the ONC), simulations and observations of massive-star feedback suggest that the gas component is undergoing rapid dispersal~\citep[e.g.][]{kroupa:2001, dinnbier-walch:2020, pabst:2019, pabst:2020}. Hence, its dynamical contribution is likely sub-dominant over the 300-Myr timescale considered here.
This section describes the methodology used to characterise our clusters prior to the simulation phase.

\subsection{Radial density profile}

\begin{figure*}
    \centering
    \includegraphics[width=0.95\textwidth]{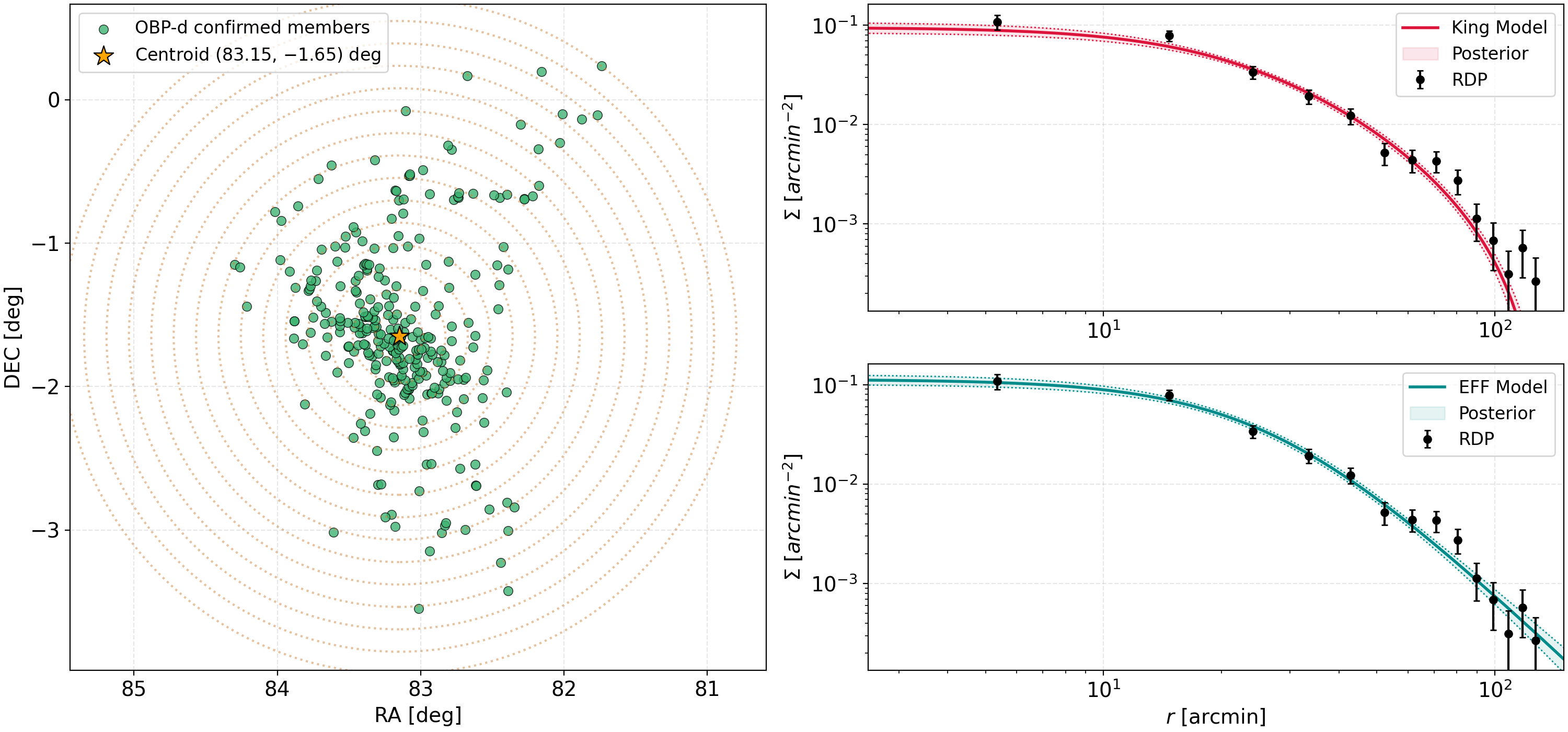}
    \caption{Fitting of the radial surface density profile for OBP-d. \textit{Left panel}: spatial distribution of members in equatorial coordinates with the concentric annuli used for star counts. \textit{Top righ}t: King-profile fit. \textit{Bottom right}: EFF-profile fit. Black points show the annular surface densities as a function of projected radius and shaded bands encloses the 16–84th percentiles of the posterior.}
    \label{fig:rdp}
\end{figure*}

To characterise a star cluster, it is necessary to determine its structural parameters such as its concentration and radial size. These can be obtained from the stellar spatial distribution through the construction of a radial density profile (RDP). The RDP is built by computing the surface number density of stars in concentric annuli around the cluster centroid, defined as the median position of the members in sky coordinates. Although this approach assumes an approximately spherical stellar distribution in the cluster potential, which is a limitation of the method, we applied it homogeneously in all clusters.
The binned surface density, $\Sigma$, is given by
\begin{equation}
    \label{rho}
    \Sigma = \frac{N_{\mathrm{bin}}}{4\pi(r_f^2 - r_i^2)},
\end{equation}
where $N_{\mathrm{bin}}$ is the number of stars in the annulus bounded by the inner radius $r_i$ and the outer radius $r_f$. Bin widths are set by the Freedman–Diaconis rule~\citep{fd:1981}. Additionally, the corresponding uncertainty, $\delta \Sigma$, can be estimated using Poisson statistics ($\delta\Sigma\propto\sqrt{N_{bin}}$).
%\begin{equation}
%    \label{rho_e}
%    \delta\Sigma = \frac{\sqrt{N_{bin}}}{4\pi(r_f^2 - r_i^2)}.
%\end{equation}

Cluster properties were derived by fitting two widely used analytical profiles to the stellar surface number density: the \citet{king:1962} profile and the Elton-Fall-Freeman (EFF) profile \citep{eff:1987}. The King profile provides a robust estimate of the cluster size through an outer cutoff parameter, $r_t$, hereafter referred to as the truncation radius. The surface density as a function of projected radius is:
\begin{equation}
    \label{king_62}
    \Sigma_k(r) = k\left[\frac{1}{\sqrt{1+(r/r_c)^2}}-\frac{1}{\sqrt{1+(r_t/r_c)^2}}\right]^2,
\end{equation}
where $r_c$ is the core radius and $k$ a normalization constant that defines the central density $\Sigma_{0,k}$ as:
\begin{equation}
    \label{k_central}
    k = \Sigma_{0,k}\left[\frac{1}{\sqrt{1+(r_t/r_c)^2}}\right]^{-2}.
\end{equation}

In parallel, the EFF profile was applied to estimate the dynamical state~\citep{portegies:2010}. Its surface density distribution is
\begin{equation}
    \label{eff_profile}
    \Sigma_e(r) = \Sigma_{0,e}\left(1 + \frac{r^2}{a^2}\right)^{-\gamma/2}.
\end{equation}
Here, $\Sigma_{0,e}$ is the central surface density, $a$ is a length scale related to the core concentration, and $\gamma$ defines the slope of the outer stellar density profile.

The fitting of the King and EFF profiles to the measured stellar surface number density was performed with the Python library \texttt{PyMC}, using Bayesian inference to obtain posterior distributions of the model parameters. Priors were defined as follows: for the central densities, in both models, we adopted a normal distribution centred on the surface density of the innermost bin, $\Sigma_o\sim \mathcal{N}(\Sigma_{o,i}, \delta\Sigma_{o,i}^2)$, with $\delta\Sigma_{o,i}$ as the dispersion of the innermost bin. In the King profile, the truncation radius was assigned a normal prior centred on the outermost radial bin, $r_t\sim \mathcal{N}(r_N, \sigma_{r,N}^2)$, where $r_N$ is the distance to the last bin and $\sigma_{r,N}$ its uncertainty associated with the bin width. The core radius was instead drawn from a uniform distribution extending from zero to the same outer limit, $r_c\sim \mathcal{U}(0, r_N)$. 

For the EFF profile, the scale parameter was assigned the same prior as the King core radius, $a \sim \mathcal{U}(0, r_N)$, and the outer slope was drawn from a uniform prior $\gamma \sim \mathcal{U}(2.0, 12.0)$, consistent with values reported for young clusters~\citep{portegies:2010,wright:2024}. As an illustration, Figure~\ref{fig:rdp} shows the fitting results for the OBP-d cluster using both models, together with the spatial distribution of sources and the concentric annuli used to calculate the surface brightness.

The method was applied to all \textit{Big Structure} groups, with three exceptions: Orion-Y (previously excluded in Section \ref{sec:data}), $\lambda$ Orionis and OBP-near. The $\lambda$ Orionis association is known to comprise three smaller substructures: Collinder 69 in the central region, and B30 together with B35 in the outskirts~\citep{barrado:2007, mathieu:2008}. 
The latter two introduce local overdensities that prevent a reliable RDP fit. We therefore restricted the analysis to Collinder 69, which contains the majority of sources. On the other hand, the OBP-near group displays a ring-like morphology with a deficit of stars around its centroid, producing large uncertainties in the central density estimates. Following \citet{sansan:2024}, where two distinct substructures were identified in OBP-near, we treated them as separate clusters, hereafter denoted as OBP-near A and OBP-near B.

From this procedure, we obtained core radii in the range $r_c = 0.16$–$1.24$ degrees and truncation radii in the range $r_t = 1.70$–$3.88$ degrees. For the EFF profile, the slope parameter was found to span $\gamma = 2.59$–$8.79$. A complete list of the fitted parameters for the 10 Big Structures and the three additional sub-structures, from both King and EFF profiles, is provided in Table~\ref{tab:rdp_fitting} in Appendix~\ref{app:rdps_params}.

\subsection{3D Velocity Distributions}

\begin{figure*}
    \centering
    \includegraphics[width=0.95\textwidth]{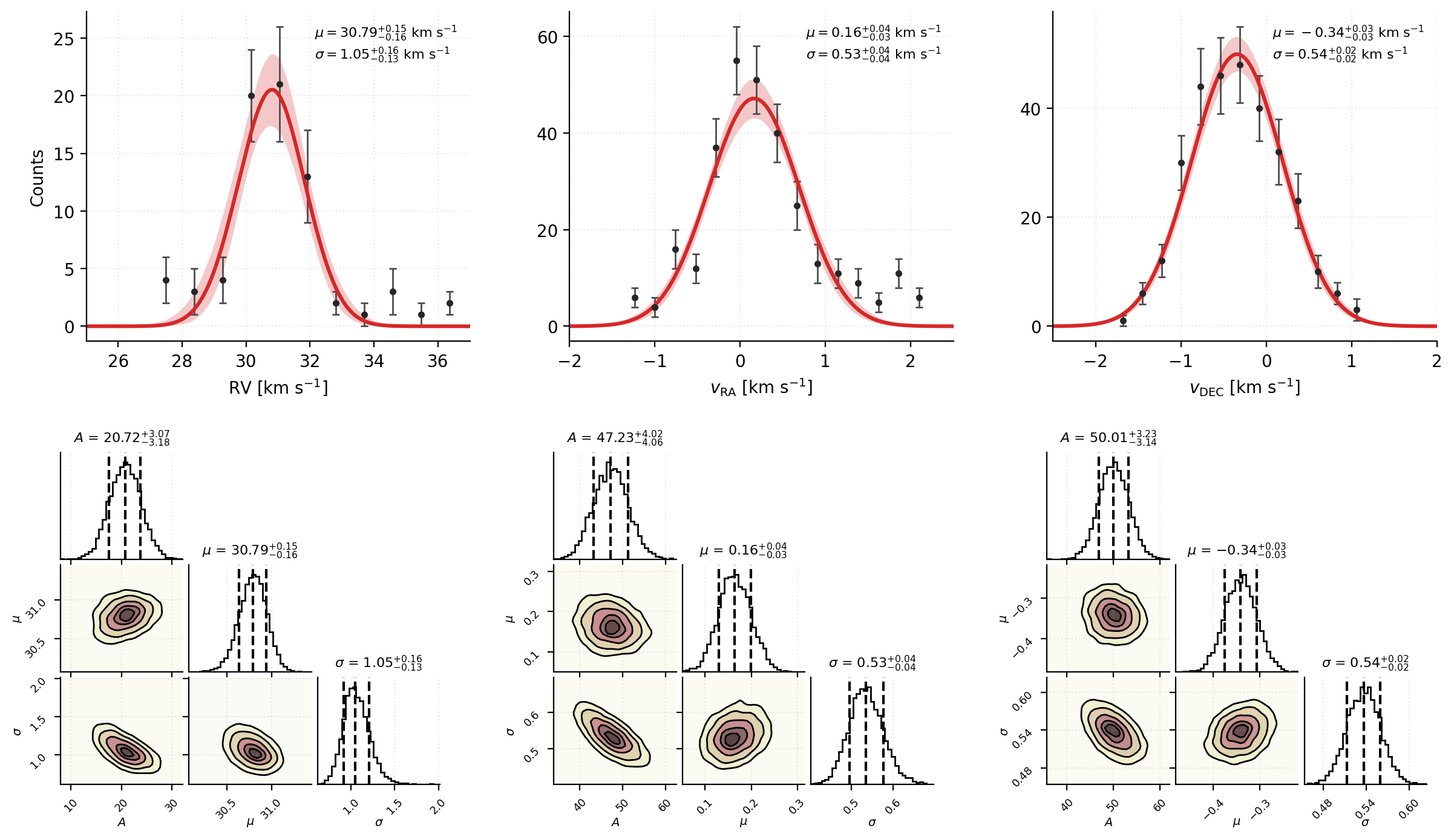}
    \caption{Example of the velocity–distribution fitting for the OBP-d cluster. \textit{Top panels}: observed histograms of RV, $V_{\mathrm{RA}}$, and $V_{\mathrm{DEC}}$. Red shaded bands encloses the 16–84th percentiles of the posteriors with the best-fit Gaussian model (red curve). On the top-right we indicate the estimation of $\mu$, and $\sigma$. \textit{Bottom panels}: posterior distributions and parameter correlations for the Gaussian amplitude ($A$), mean velocity ($\mu$), and dispersion ($\sigma$) obtained with \texttt{PyMC}.}
    \label{fig:velocities}
\end{figure*}

To analyse the 3D kinematics of the OSFC clusters, we combined proper motions and distances with radial velocities to reconstruct the distribution in the velocity space. We modelled each component of the distribution with a trivariate Gaussian aligned with the observed coordinate system, allowing us to estimate velocity dispersions in all directions. The analytical form is
\begin{equation}
    \label{vel_gauss}
    f(v) = A \exp\left[-0.5\left(\frac{v-\mu}{\sigma_v}\right)^2\right],
\end{equation}
where $A$ is the Gaussian amplitude, $\mu$ the mean velocity, and $\sigma_v$ the velocity dispersion.

The fittings were performed independently for RV, $V_{\mathrm{RA}}$, and $V_{\mathrm{DEC}}$ using \texttt{PyMC}. Note that the size of the RV sample in each structure is smaller than that for $V_{\mathrm{RA}}$ and $V_{\mathrm{DEC}}$, as mentioned in Section \ref{sec:data}. The velocities on the sky ($V_{\mathrm{RA}}$, and $V_{\mathrm{DEC}}$) are estimated from the proper motions and parallaxes of the members of the stellar clusters\footnote{Calculated as $V_{RA}=4.74\mu_{\alpha}^*/\varpi$ and $V_{DEC}=4.74\mu_{\delta}/\varpi$}.
The observed histograms, with bin widths set by the Freedman–Diaconis rule, served as likelihoods. We adopted the following priors: $A \sim \mathcal{N}(\max(A), \sigma_A^2)$, where $\max(A)$ is the most populated bin and $\sigma_A$ its Poisson uncertainty; $\mu \sim \mathcal{U}(v_{\min}, v_{\max})$, with $v_{\min}$ and $v_{\max}$ given by the data range; and $\sigma_v \sim \mathcal{U}(0.5, 10.0)$, consistent with velocity dispersions reported for YSCs~\citep{wright:2024}. As an example, Figure~\ref{fig:velocities} shows the fitting process for the OBP-d cluster together with the posterior distributions of the three parameters in each velocity component. Finally, the total velocity dispersion was derived from the posterior samples as $\sigma_T^2 = \sigma_{RV}^2 + \sigma_{RA}^2 + \sigma_{DEC}^2$, with uncertainties propagated through Monte Carlo (MC) sampling.

The velocity distribution fits for the OSFC clusters are presented in Appendix~\ref{app:velocity}, where Table~\ref{tab:velocity_fit} lists the fitted parameters. As a result, $\sigma_T$ spans 1.02–4.06 km s$^{-1}$ in our clusters. This range is consistent with previous studies of the OSFC and other OB associations, with typical dispersions of 1–5 km s$^{-1}$~\citep[e.g.][]{tobin:2009, daRio:2017, wright:2024}. The lower values ($\sim$1 km s$^{-1}$) correspond to dynamically cold systems that may remain bound, whereas higher values ($\sim$4 km s$^{-1}$) suggest unbound clusters undergoing dispersion~\citep{wright_mamajek:2018, kounkel_covey:2019}.

\subsection{Age and Mass Estimation}

To estimate the ages of the \textit{Big Structures}, we fit isochrones to distance-corrected colour–magnitude diagrams (CMDs) using \textit{Gaia} photometry and the PAdova and TRieste Stellar Evolution Code \citep[\textsc{parsec};][]{marigo:2017}. Distances were derived from parallaxes via $d = 1/\varpi$. Following similar studies~\citep{zari:2019, swiggum:2021}, the method employs standard likelihood maximization, assuming Gaussian uncertainties in all observed quantities and adopting age, $\tau$, and extinction, $A_V$, as free parameters. The log-likelihood for a single star of mass $m$ to belong to an isochrone with parameters $\theta = (\tau, A_V)$ is expressed as:
\begin{equation}
    \label{likelihood}
    \ln (L(\theta, m)) = \ln \left(\frac{1}{(2\pi)^{1/2}\sigma_G}\right)+\ln \left(\frac{1}{(2\pi)^{1/2}\sigma_{BR}}\right)-\frac{\chi^2}{2}.
\end{equation}
with,
\begin{equation}
    \label{chi2}
    \chi^2 = \left(\frac{M_G^{obs} - M_G(\theta, m)}{\sigma_G}\right)^2 + \left(\frac{C_{BR}^{obs} - C_{BR}(\theta, m)}{\sigma_{BR}}\right)^2
\end{equation}
where $M_G$ is the absolute magnitude in the \textit{Gaia} $G$ band, and $C_{BR}$ is the colour index $B_P - R_P$, with respective uncertainties $\sigma_M$ and $\sigma_{BR}$. The superscript “\textit{obs}” denotes observed quantities. Since stars are not uniformly distributed along the CMD locus, we applied a normalized weight to the $j$-th likelihood using a factor $w$ computed with the \texttt{binned\_statistic\_2d} function from the \texttt{SciPy} library~\citep{scipy}, used to calculate 2D binned statistics. This weighting scheme increases the relative contribution of the low-mass end of the stellar population, where the number of sources is larger compared to their high-mass counterparts. Therefore, the log-likelihood for the $N$ members of the cluster is defined as:
\begin{equation}
    \label{L_comb}
    \ln (L_{comb}(\theta, m)) = \sum_{j=i}^N w_j\ln(L_j(\theta, m))
\end{equation}
For this analysis, we employed a grid of models spanning ages from $1.0$ to $50.0$ Myr in steps of $0.05$ Myr, and extinction values from $0.0$ to $1.0$ mag in steps of $0.05$ mag. A fixed solar metallicity of $Z = Z_\odot = 0.0158$ was assumed for all models~\citep{zari:2019}. The fitting procedure does not account for the presence of unresolved binaries, photometric variability in young stars, circumstellar material, or possible age spreads within individual groups. We estimated $\tau$ and $A_V$ for all \textit{Big Structures} by minimizing Equation~\eqref{L_comb}. Uncertainties were estimated from 100 MC realizations, in which 20\% of the data were resampled and observational uncertainties were propagated assuming Gaussian errors. We report parameter uncertainties as the 16th--84th percentile ranges of the resulting distributions.

\begin{figure}
    \centering
    \includegraphics[width=0.47\textwidth]{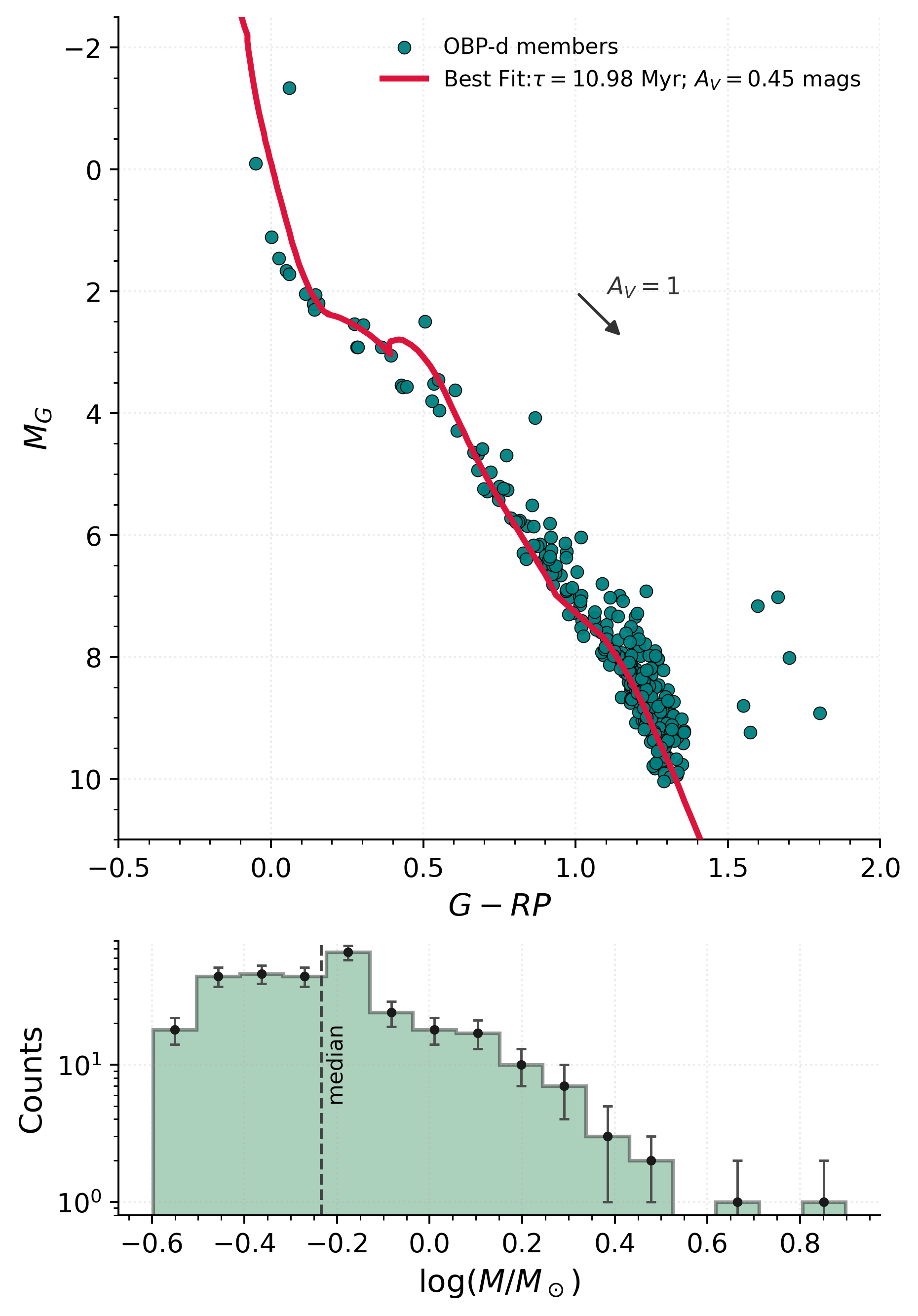}
    \caption{Isochrone fitting for the OBP-d cluster. \textit{Top panel}: \textit{Gaia} colour–magnitude diagram with the observed members (green points) and the best-fitting isochrone (red line). The arrow denotes the extinction vector. \textit{Bottom panel}: resulting stellar mass distribution derived from interpolation along the best-fitting isochrone.}
    \label{fig:age_mass}
\end{figure}

To estimate stellar masses, we first computed the minimum distance between each star and the best-fitting isochrone, increasing the resolution of the model in both colour and absolute magnitude using \texttt{SmoothBivariateSpline} interpolation from the \texttt{SciPy} library. To account for uncertainties, we applied a MC approach by sampling the colour and magnitude of each star. Figure~\ref{fig:age_mass} illustrates this process for the OBP-d cluster, showing both the isochrone fit and the resulting stellar mass distribution. The total observed mass of the cluster, $M_{\mathrm{obs}}$, was obtained by summing the individual stellar contributions. Isochrone fits and corresponding mass distributions for the remaining groups are presented in Appendix~\ref{app:iso_fit}.

An important caveat of our method applies to highly embedded clusters such as the ONC. In these systems, the broad pre–main-sequence locus in the CMD can bias isochrone fits toward older ages and lower $A_V$ \citep[e.g.][]{daRio:2016}. Underestimation of $A_V$ propagates into the inferred stellar masses and therefore the total cluster mass, so the reported values should be treated with caution.

\subsection{Estimating the Virial State}

The virial state, $\alpha_{vir}$, quantifies the present gravitational boundedness of a system. In its classical form, it is defined from the energy balance as $\alpha_{vir}\equiv 2T/|W|$, where $T$ and $W$ are the total kinetic and gravitational potential energies, respectively. In practice, direct estimates of $T$ and $W$ require complete 6D phase-space and a well-constrained mass distribution, which are not available for all OSFC clusters. We therefore adopt an observationally equivalent expression, comparing the measured total dispersion $\sigma_{T,o}$ to the dispersion expected for virial equilibrium, $\sigma_{\mathrm{vir}}$, in the observed stellar component~\citep{portegies:2010, wright:2024}:
\begin{equation}
\label{virial}
\alpha_{vir} = \frac{\sigma_{T,o}}{\sigma_{\mathrm{vir}}},
\end{equation}
with $\alpha_{\mathrm{vir}} < 1$ for a sub-virial system, $\alpha_{\mathrm{vir}} = 1$ for virial equilibrium, and $\alpha_{\mathrm{vir}} > 1$ for a super-virial state. $\sigma_{\mathrm{vir}}$ is computed from the virial theorem expressed as:
\begin{equation}
\label{sigma_vir_3D}
\sigma_{3D,\mathrm{vir}} = \sqrt{\frac{G M_{cl}}{2 r_{\mathrm{vir}}}},
\end{equation}
where $G$ is the gravitational constant, $M_{cl}$ the total cluster mass, and $r_{\mathrm{vir}}$ the virial radius. By introducing the parameter $\eta = 6 r_{\mathrm{vir}} / r_{\mathrm{eff}}$~\citep{portegies:2010}, with $r_{\mathrm{eff}}$ as the projection of the effective (half-light) radius, this relation becomes
\begin{equation}
\label{sigma_vir_eta}
\sigma_{3D,\mathrm{vir}} = \sqrt{\frac{3 G M_{\mathrm{cl}}}{\eta r_{\mathrm{eff}}}}.
\end{equation}
The value of $\eta$ can be derived from the $\gamma$ parameter in the EFF profile using the relation given by \citet{portegies:2010}. Consequently, we measure $r_{\mathrm{eff}}$ under the assumption that all stars have equal luminosity.

Table \ref{tab:full_params} summarizes the structural and dynamical parameters derived for the OSFC clusters. The effective radii range from small compact groups such as $\sigma$ Ori ($r_{\rm eff}\sim2$ pc) to more extended associations like Ori-North and OBP-Far ($r_{\rm eff}>6$ pc), consistent with the diversity of morphologies reported in previous studies of the Orion complex \citep[e.g.][]{kuhn:2019}. The inferred $r_{\mathrm{vir}}$ and $\eta$ parameters show that all systems seem to be observationally unbound, in agreement with expectations for young, substructured populations. The observed three-dimensional velocity dispersions cover the range $\sim$1–4 km s$^{-1}$, comparable to those reported for other nearby OB associations \citep{wright:2024}.

Comparison between the observed dispersions and the virial values reveals that all clusters are supervirial. In particular, several OBP subgroups and Ori-East reach extreme value $\alpha_{\rm vir}\gtrsim 15$, suggesting that they are rapidly expanding. By contrast, denser systems such as the ONC and Briceño clusters have moderate $\alpha_{\rm vir}$ values ($\sim$4–5), consistent with slower expansion and a greater capacity to retain a bound remnant. This overall picture supports the view that the OSFC spans a range of supervirial dynamical states, likely imprinted by varying initial conditions and early gas expulsion. We note that these estimates are based on the observed component alone and may be biased by incompleteness in the membership census. We address this effect in Section~\ref{sec:pdmf}.

%%%%%%%%%%
%%%%%%%%%%
\begin{landscape}
{\renewcommand{\arraystretch}{1.4}
\begin{table}
\centering
\caption{Structural and dynamical parameters of the OSFC clusters. Columns are: (1) cluster name; (2) core radius $r_c$; (3) truncation radius $r_t$; (4) effective (half-light) radius $r_{\mathrm{eff}}$; (5) virial radius $r_v$; (6) $\eta$ parameter derived from the Elson–Fall–Freeman (EFF) profile fits, linked to the slope $\gamma$ of the outer density profile; (7) total observed 3D velocity dispersion $\sigma_T$; (8) cluster age $\tau$ from isochrone fitting; (9) visual extinction $A_V$; (10) stellar mass $M_{cl}$ derived from confirmed members; (11) expected virial velocity dispersion $\sigma_{vir}$ assuming the observed mass is in virial equilibrium; and (12) virial parameter $\alpha_{vir} = \sigma_T / \sigma_{vir}$, quantifying the dynamical state of the system. Errors reflect propagated observational uncertainties and posterior sampling of the fitted parameters.}
\label{tab:full_params}
\begin{tabular}{lccccccccccc}
\hline
\multicolumn{1}{c}{Cluster}  & \begin{tabular}[c]{@{}c@{}}$r_c$\\ (pc)\end{tabular} & \begin{tabular}[c]{@{}c@{}}$r_t$\\ (pc)\end{tabular} & \begin{tabular}[c]{@{}c@{}}$r_{eff}$\\ (pc)\end{tabular} & \begin{tabular}[c]{@{}c@{}}$r_v$\\ (pc)\end{tabular} & $\eta$                  & \begin{tabular}[c]{@{}c@{}}$\sigma_{T}$\\ (km s$^{-1}$)\end{tabular} & \begin{tabular}[c]{@{}c@{}}$\tau$\\ (Myr)\end{tabular} & $A_V$                  & \begin{tabular}[c]{@{}c@{}}$M_{cl}$\\ ($M_\odot$)\end{tabular} & \begin{tabular}[c]{@{}c@{}}$\sigma_{vir}$\\ (km s$^{-1}$)\end{tabular} & $\alpha_{vir}$ \\
\multicolumn{1}{c}{(1)}      & (2)                                                  & (3)                                                  & (4)                                                      & (5)                                                  & (6)                     & (7)                                                                  & (8)                                                   & (9)                   & (10)                                                           & (11)                                                                   & (12)           \\ \hline
$\lambda$ Ori (Collinder 69) & 4.97$\pm$1.35                                        & 16.75$\pm$0.78                                       & 4.39$\pm$0.35                                            & 7.78$\pm$0.74                                        & 10.67$^{+0.22}_{-0.26}$ & 1.36$^{+0.11}_{-0.10}$                                               & 6.82$^{+0.84}_{-1.45}$                                 & 0.35$^{+0.05}_{-0.02}$ & 274.41                                                         & 0.27$\pm$0.1                                                           & 4.98$\pm$0.39  \\
Ori-North                    & 8.17$\pm$1.24                                        & 25.47$\pm$2.34                                       & 8.53$\pm$1.14                                            & 14.66$\pm$2.27                                       & 10.37$^{+0.40}_{-0.36}$ & 2.51$^{+0.30}_{-0.29}$                                               & 18.10$^{+1.67}_{-0.96}$                                & 0.25$^{+0.05}_{-0.05}$ & 237.06                                                         & 0.19$\pm$0.1                                                           & 13.46$\pm$1.96 \\
Briceño-1A                   & 3.19$\pm$0.45                                        & 12.01$\pm$0.57                                       & 3.33$\pm$0.28                                            & 5.51$\pm$0.59                                        & 9.92$^{+0.42}_{-0.32}$  & 1.05$^{+0.15}_{-0.14}$                                               & 11.01$^{+0.76}_{-0.25}$                                & 0.30$^{+0.05}_{-0.08}$ & 140.72                                                         & 0.23$\pm$0.1                                                           & 4.48$\pm$0.68  \\
Briceño-1B                   & 1.68$\pm$0.13                                        & 15.01$\pm$0.64                                       & 2.59$\pm$0.18                                            & 4.43$\pm$0.39                                        & 10.27$^{+0.24}_{-0.23}$ & 1.02$^{+0.11}_{-0.10}$                                               & 12.39$^{+1.24}_{-1.01}$                                & 0.33$^{+0.10}_{-0.10}$ & 163.74                                                         & 0.28$\pm$0.1                                                           & 3.63$\pm$0.43  \\
Ori-East                     & 1.54$\pm$0.18                                        & 12.75$\pm$0.74                                       & 4.43$\pm$2.14                                            & 7.99$\pm$1.70                                        & 10.73$^{+0.27}_{-0.26}$ & 2.63$^{+0.26}_{-0.25}$                                               & 3.10$^{+1.05}_{-0.30}$                                 & 0.05$^{+0.10}_{-0.05}$ & 69.85                                                          & 0.14$\pm$0.3                                                           & 19.6$\pm$5.51  \\
OBP-Far                      & 3.81$\pm$0.32                                        & 37.52$\pm$1.23                                       & 6.19$\pm$0.45                                            & 10.41$\pm$0.95                                       & 10.14$^{+0.32}_{-0.31}$ & 3.48$^{+0.42}_{-0.41}$                                               & 15.72$^{+1.05}_{-0.30}$                                & 0.45$^{+0.05}_{-0.05}$ & 178.97                                                         & 0.19$\pm$0.1                                                           & 18.15$\pm$2.37 \\
$\sigma$ Ori                 & 1.07$\pm$0.09                                        & 12.01$\pm$0.73                                       & 1.86$\pm$0.17                                            & 3.21$\pm$0.37                                        & 10.39$^{+0.30}_{-0.30}$ & 2.04$^{+0.19}_{-0.18}$                                               & 3.25$^{+0.46}_{-1.07}$                                 & 0.40$^{+0.10}_{-0.15}$ & 144.36                                                         & 0.31$\pm$0.2                                                           & 6.68$\pm$0.74  \\
OBP-b                        & 4.97$\pm$0.57                                        & 17.03$\pm$0.98                                       & 4.61$\pm$0.29                                            & 7.20$\pm$0.47                                        & 9.36$^{+0.13}_{-0.08}$  & 4.06$^{+0.57}_{-0.56}$                                               & 18.89$^{+0.63}_{-0.49}$                                & 0.32$^{+0.08}_{-0.07}$ & 163.80                                                         & 0.22$\pm$0.1                                                           & 18.24$\pm$2.62 \\
OBP-d                        & 2.64$\pm$0.25                                        & 16.14$\pm$0.75                                       & 3.58$\pm$0.24                                            & 6.13$\pm$0.49                                        & 10.31$^{+0.23}_{-0.22}$ & 1.30$^{+0.12}_{-0.11}$                                               & 10.98$^{+0.95}_{-0.89}$                                & 0.50$^{+0.05}_{-0.03}$ & 220.24                                                         & 0.28$\pm$0.1                                                           & 4.66$\pm$0.39  \\
OBP-Near A                   & 2.40$\pm$0.43                                        & 10.43$\pm$0.64                                       & 2.93$\pm$0.43                                            & 5.01$\pm$0.98                                        & 10.32$^{+0.57}_{-0.60}$ & 2.75$^{+1.28}_{-1.27}$                                               & 10.72$^{+1.17}_{-0.92}$                                & 0.20$^{+0.08}_{-0.10}$ & 89.03                                                          & 0.20$\pm$0.2                                                           & 14.30$\pm$6.94 \\
OBP-Near B                   & 5.28$\pm$0.62                                        & 12.37$\pm$0.52                                       & 4.35$\pm$0.25                                            & 6.79$\pm$0.42                                        & 9.37$^{+0.15}_{-0.09}$  & 3.02$^{+0.61}_{-0.60}$                                               & 6.71$^{+0.50}_{-1.15}$                                 & 0.25$^{+0.05}_{-0.01}$ & 114.45                                                         & 0.19$\pm$0.1                                                           & 16.12$\pm$3.26 \\
ONC                          & 3.32$\pm$0.20                                        & 17.18$\pm$0.67                                       & 3.99$\pm$0.15                                            & 6.34$\pm$0.23                                        & 9.54$^{+0.07}_{-0.07}$  & 2.98$^{+0.07}_{-0.06}$                                               & 4.08$^{+0.56}_{-2.10}$                                 & 0.50$^{+0.05}_{-0.05}$ & 1230.11                                                        & 0.64$\pm$0.1                                                           & 4.65$\pm$0.14  \\
Ori-South                    & 6.41$\pm$0.89                                        & 16.67$\pm$0.77                                       & 4.99$\pm$0.28                                            & 7.87$\pm$0.43                                        & 9.46$^{+0.17}_{-0.14}$  & 2.12$^{+0.12}_{-0.11}$                                               & 3.63$^{+0.62}_{-0.53}$                                 & 0.50$^{+0.05}_{-0.06}$ & 187.09                                                         & 0.22$\pm$0.1                                                           & 9.57$\pm$0.62  \\ \hline
\end{tabular}
\end{table}}
\end{landscape}

%%%%%%%%%%
%%%%%%%%%%

\begin{figure}
    \centering
    \includegraphics[width=0.48\textwidth]{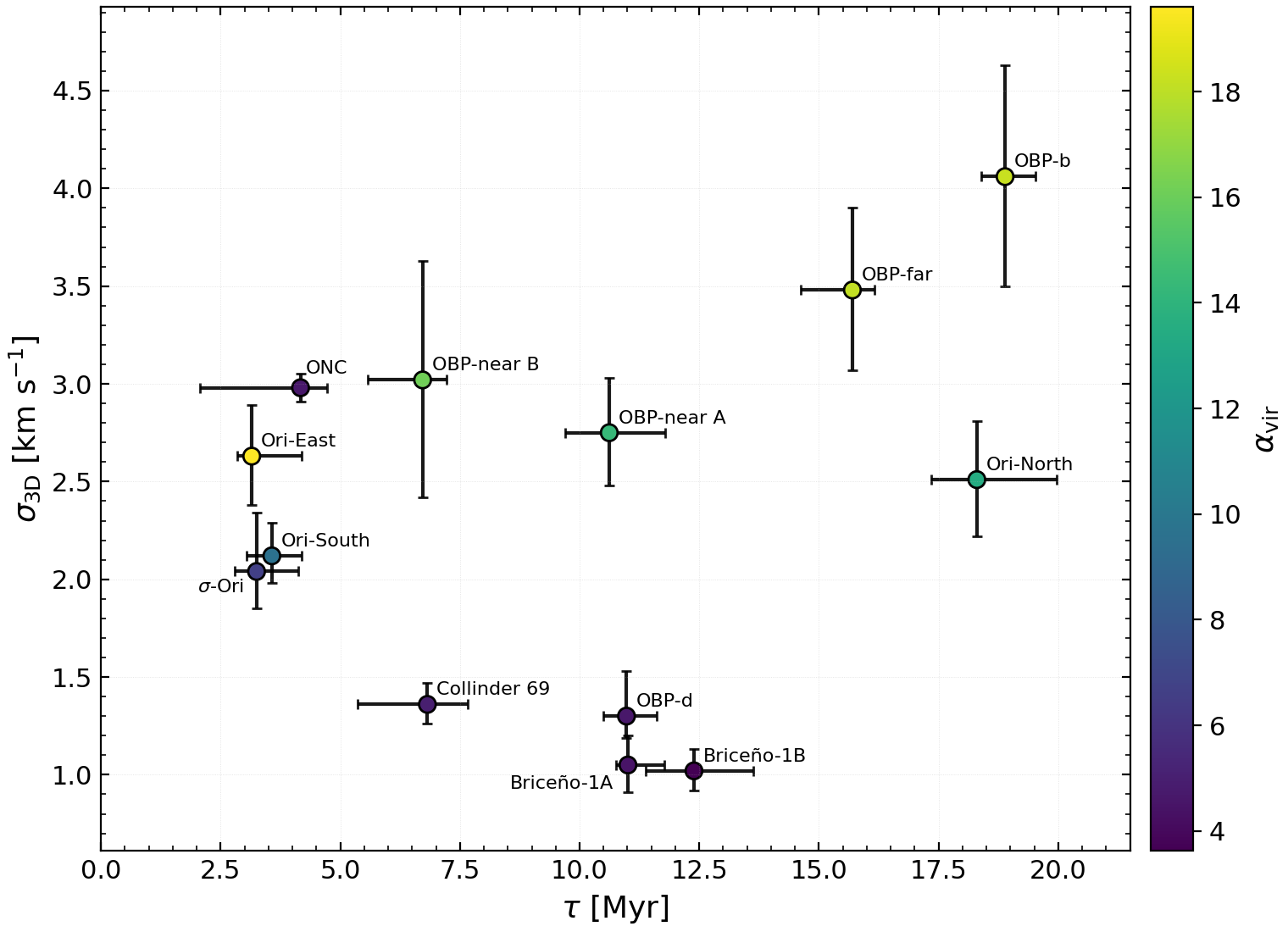}
    \caption{Three-dimensional velocity dispersion, $\sigma_{3D}$, as a function of cluster age, $\tau$, for the OSFC. The colour scale indicates the virial parameter, $\alpha_{vir}$. The uncertainties correspond to the 16th--84th percentiles.}
    \label{fig:sigma_tau}
\end{figure}

%%%%%%
%%%%%%
%%%%%

%In Figure~\ref{fig:sigma_tau}, we show the relation between $\sigma_{3D}$ and $\tau$ for the OSFC clusters, with circle colours encoding the observed $\alpha_{\rm vir}$. Three features stand out. At $\tau\lesssim 5$ Myr, clusters gather around $\sigma_{3D}\simeq 2$–$3~\mathrm{km,s^{-1}}$, consistent with velocities set by the natal potential~\citep{portegies:2010}. For $\tau\gtrsim 5$ Myr, the population separates into a cold branch with low-to-moderate $\alpha_{\rm vir}$ and $\sigma_{3D}\approx 1~\mathrm{km,s^{-1}}$, comparable to the low dispersions reported for nearby young clusters~\citep{wright:2024}, and a hot branch with $\sigma_{3D}\gtrsim 2~\mathrm{km,s^{-1}}$ and systematically larger $\alpha_{\rm vir}$. The oldest groups in our sample ($\tau>15$ Myr) lie predominantly on the hot branch.
%%%%%
%%%%%

In Figure~\ref{fig:sigma_tau}, we show the relationship between $\sigma_{3D}$ and $\tau$ for the OSFC clusters, with circle colours encoding the observed $\alpha_{\rm vir}$. Two features stand out. At $\tau\lesssim 5$ Myr, clusters gather around $\sigma_{3D}\simeq 2$–$3~\mathrm{km,s^{-1}}$, consistent with velocities set by the natal potential~\citep{portegies:2010}. For $\tau\gtrsim 5$ Myr, the sample suggests a separation into two kinematic regimes: a cold population with low-to-moderate $\alpha_{\rm vir}$ and $\sigma_{3D}< 2~\mathrm{km\ s^{-1}}$, comparable to the low dispersions reported for nearby young clusters~\citep{wright:2024}, and an apparent hot population with $\sigma_{3D}> 2~\mathrm{km\ s^{-1}}$ and systematically larger $\bmath{\alpha_{\rm vir}}$.

A plausible evolutionary route for clusters on the cold regime is gradual, quasi-adiabatic gas removal: when the gas-expulsion timescale exceeds the dynamical time, the stellar system expands slowly while preserving a bound core, yielding low dispersions and low-to-moderate $\alpha_{\rm vir}$~\citep{hills:1980, boily:2003}. Rapid gas removal, by contrast, leaves the stars strongly supervirial and drives fast expansion, populating the hot regime~\citep{baumgardt:2007}. Because expansion alone tends to cool self-gravitating systems, maintaining high dispersions likely requires continued heating from external perturbations and possibly internal sources such as binary heating in collisional systems~\citep{heggiehut:2003}. Quantifying the effects of binarity is beyond the scope of this work and is therefore deferred to future work (Sánchez-Sanjuán in prep.).

More generally, the bifurcation might reflect time-variable formation conditions within the OSFC, producing colder and hotter clusters. Taken together, these results suggest that cluster kinematics in Orion depend sensitively on both initial conditions and subsequent environmental processing, yielding a scattered distribution in $\sigma_{3D}$ rather than a simple age–dispersion correlation.

%%%%%%
%%%%%%
%%%%%%

\section{Completeness of the Mass Function}
\label{sec:pdmf}

\subsection{Completeness Determination}

Reliable identification of cluster members in the OSFC is affected by both methodological filters (e.g., cuts in RUWE, fidelity, and parallax error) and \textit{Gaia}’s instrumental limitations (e.g., faint-end sensitivity and bright-star saturation). These biases result in a truncation of the observed present-day mass function (PDMF). We therefore estimate for each cluster the mass completeness ratio and apply it to the selected mass function to recover the underlying PDMF.

\begin{figure}
    \centering
    \includegraphics[width=0.46\textwidth]{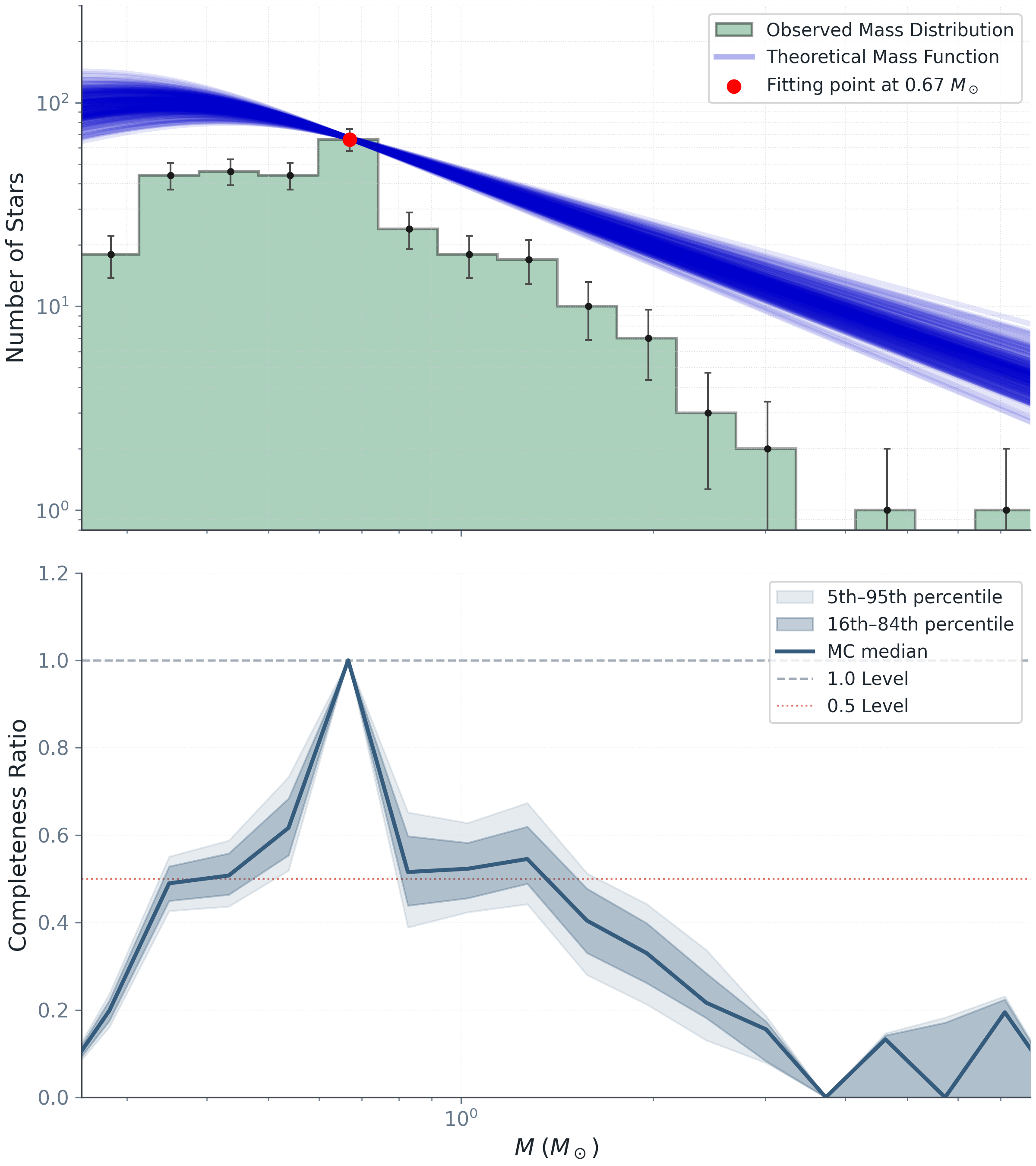}
    \caption{Completeness correction procedure applied to the observed PDMF of the OBP-d cluster. \textit{Top panel}: Observed stellar mass distribution (green histogram) compared with the tapered power-law model anchored at the most populated mass bin (red point at $\sim 0.67 M_\odot$). Blue curves show 300 realizations of the PDMF. \textit{Bottom panel}: Completeness ratio trend as a function of stellar mass obtained from 1000 MC realizations including stellar mass uncertainties; shaded regions indicate the 16th--84th and 5th--95th percentile ranges.}
    \label{fig:mass_function}
\end{figure}

To reconstruct the intrinsic PDMF, we assume that the peak of the observed mass histogram marks the onset of incompleteness. For all OSFC clusters, this peak lies between $0.3$–$0.7M_\odot$, a range where \textit{Gaia} photometry is both unsaturated and highly complete, in agreement with assessments of survey sensitivity~\citep[e.g.][]{Riello:2021, Fabricius:2021, cantat-gaudin:2023}. To model the underlying distribution, we used a tapered power-law function, following~\citet{suarez:2019}, which reproduces the canonical power-law behaviour at intermediate-to-high masses while introducing an exponential taper at low masses to capture the observed turnover. The function is defined as
\begin{equation}
\label{tappered_mf}
\epsilon (\log m) \propto m^{-\Gamma}\left[1 - e^{-(m/m_p)^\beta}\right],
\end{equation}
where $\Gamma$ is the high-mass slope, $\beta$ the tapering exponent, and $m_p$ the characteristic turnover mass. For the 25Ori region, \citet{suarez:2019} found $\Gamma = 1.10 \pm 0.09$, $\beta = 2.11 \pm 0.09$, and $m_p = 0.31 \pm 0.03$ M$_\odot$ by combining optical and near-IR photometry of 1,687 candidate members. These values are consistent with other determinations of the low-mass turnover~\citep[e.g.][]{chabrier:2003, boudreault:2012}. Assuming that OSFC clusters share a comparable PDMF and that the membership determination in the \citet{sansan:2024} catalogue is reliable, we anchored the adopted PDMF at the most populated mass bin.  This choice provides a robust normalisation point while minimising the risk of missing high-probability members (typically >90\% from the HDBSCAN algorithm). From this model, we calculated a completeness ratio for each bin, defined as the observed number of stars relative to the expected value from the theoretical PDMF, thereby quantifying the missing fraction required to reconstruct the intrinsic distribution. Figure~\ref{fig:mass_function} illustrates the procedure applied to the OBP-d cluster.

Uncertainties were quantified through a MC approach. We repeated the procedure 300 times, keeping the observed sample fixed for each cluster. In each iteration, we drew a Poisson realization of the binned star counts while sampling the PDMF parameters within their uncertainties. This produced robust estimates of the corrected stellar number, $N_{s,\mathrm{cor}}$, and total mass, $M_{cl,\mathrm{cor}}$, together with confidence intervals. The assumption of a common PDMF is further supported by \citet{pang:2024}, who find that clusters younger than $\sim$200 Myr the PDMF slope in the solar neighbourhood remains stable.

\subsection{RDP and Dynamical Corrections}

\begin{figure*}
    \centering
    \includegraphics[width=0.98\textwidth]{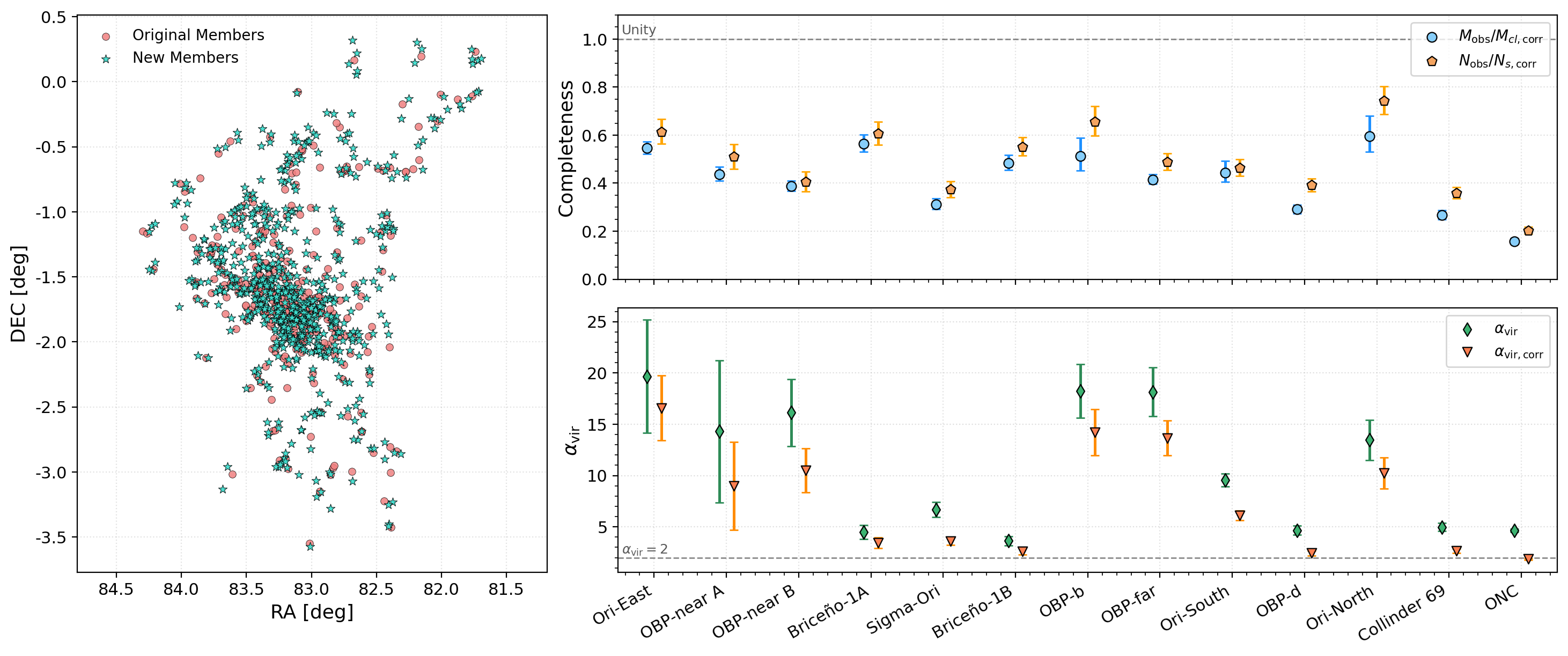}
    \caption{Impact of completeness corrections on membership, mass, and virial state. \textit{Left panel}: Sky distribution for OBP-d showing original members (salmon circles) and additional members recovered with the KDE–based reconstruction (cyan stars). \textit{Right–top}: Completeness factors $M_{\rm obs}/M_{\rm cor}$ (blue circles) and $N_{\rm obs}/N_{\rm cor}$ (orange symbols). \textit{Right–bottom}: Virial parameter before (green diamonds) and after (orange triangles) the completeness corrections. Groups are ordered by increasing observed mass and error bars show the 16th–84th percentiles from the Monte-Carlo error propagation.}
    \label{fig:mass_completeness}
\end{figure*}

The completeness analysis provides, for each cluster, the number of missing stars per mass bin as implied by the adopted PDMF. These stars must be spatially placed before refitting the structural models and deriving dynamical quantities. We therefore reconstruct the three-dimensional spatial probability density of each substructure using a Gaussian kernel density estimator (KDE) in the observed coordinates $(\alpha,\delta,\varpi)$. The KDE preserves the morphology recovered by \texttt{HDBSCAN}, assigning higher probability to the denser observed regions, where incompleteness is expected to be strong.

The KDE bandwidth is determined through an iterative calibration procedure: we generate synthetic clusters with the same size as the observed sample and vary the bandwidth until the residuals between the synthetic and observed spatial distributions are minimized. With the bandwidth fixed, the missing stars are sampled from the KDE and added to the total catalogue, assuming that there is no mass segregation within the clusters. With the corrected amount of stars per cluster, we then recompute the RDP and refit both King and EFF profiles to obtain corrected structural parameters and uncertainties. Table~\ref{tab:params_corr} lists, in columns 2 and 3, the completeness-corrected core radius $r_{c,cor}$, and truncation radius $r_{t,cor}$, respectively, from which we derive the concentration parameter $c\approx \log (r_{t,cor}/r_{c,cor})$ in column 4. Additionally, we estimated the completeness-corrected virial radius $r_{v,cor}$ in column 5 after using the relation $r_{vir}=\eta r_{eff}/6$ from~\citet{portegies:2010}.

The left panel of Figure~\ref{fig:mass_completeness} (shown for OBP--d) illustrates the effect of the completeness correction on the sky distribution: newly added members (cyan) are concentrated toward the crowded centre while preserving the large–scale morphology (filaments/elongations), confirming that the KDE does not impose an artificial geometry. The right–\emph{top} panel reports the completeness ratios for each group, expressed as $M_{\rm obs}/M_{cl,cor}$ (circles) and $N_{\rm obs}/N_{s,cor}$ (pentagons), where unity indicates full completeness. Most substructures lie in the $0.3$–$0.7$ range (with a median incompleteness of $\sim$45\%), with the ONC showing the lowest values ($\sim0.2$) owing to severe crowding and saturation in \textit{Gaia}. The completeness-corrected $N_{s,cor}$ and $M_{cl,cor}$ are listed in columns 6 and 7 of Table~\ref{tab:params_corr}

Finally, the mass correction has measurable dynamical consequences. The right–\emph{bottom} panel shows a systematic decrease in the virial parameter after correction. Across the sample, we found an average variation of $\alpha_{\rm vir,cor}/\alpha_{\rm vir} \approx 0.66$, indicating that clusters would appear up to $\sim$34\% more supervirial than they truly are. Physically, adding the missing members raises the cluster mass and thus the virial dispersion $\sigma_{\rm vir}$. Therefore, for a given observed dispersion $\sigma_{T,o}$, the ratio $\alpha_{\rm vir}=\sigma_{T,o}/\sigma_{\rm vir}$ decreases. The completeness-corrected $\alpha_{\rm vir}$ values are reported in column 8 of Table~\ref{tab:params_corr}.

{\renewcommand{\arraystretch}{1.3}
\begin{table*}
\centering
\caption{Completeness–corrected structural and dynamical parameters for the OSFC. Columns: (1) cluster name; (2) core radius $r_{c,{\rm cor}}$; (3) truncation radius $r_{t,{\rm cor}}$; (4) the concentration parameter $c\approx \log (r_{t,cor}/r_{c,cor})$; (5) virial radius $r_{v,{\rm cor}}$; (6) number of stars $N_{s,{\rm cor}}$; (7) total stellar mass $M_{cl,{cor}}$; and (8) virial ratio $\alpha_{{\rm vir},{cor}}$. Uncertainties are 1$-\sigma$ from the Monte–Carlo error propagation of counting noise and PDMF parameters.}
\label{tab:params_corr}
\begin{tabular}{lccccccc}
\hline
\multicolumn{1}{c}{Cluster}  & \begin{tabular}[c]{@{}c@{}}$r_{c,cor}$\\ (pc)\end{tabular} & \begin{tabular}[c]{@{}c@{}}$r_{t,cor}$\\ (pc)\end{tabular} & $c$ & \begin{tabular}[c]{@{}c@{}}$r_{v,cor}$\\ (pc)\end{tabular} & $N_{s,cor}$ & \begin{tabular}[c]{@{}c@{}}$M_{cl,cor}$\\ ($M_\odot$)\end{tabular} & $\alpha_{vir,cor}$ \\
\multicolumn{1}{c}{(1)}      & (2)     & (3)      & (4)     & (5)     & (6)      & (7)    & (8)                 \\ \hline \hline
$\lambda$ Ori (Collinder 69) & 3.69$\pm$0.53     & 18.91$\pm$0.75      & 0.71$\pm$0.07   & 8.20$\pm$0.56     & 1103$\pm$50  & 1029.37$\pm$73.39     & 2.63$\pm$0.21       \\
Ori-North                    & 10.40$\pm$1.32    & 27.95$\pm$2.52      & 0.43$\pm$0.04  & 14.10$\pm$1.43    & 411$\pm$22    & 398.62$\pm$50.06        & 10.24$\pm$1.50      \\
Briceño-1A                   & 3.20$\pm$0.33     & 12.29$\pm$0.52      & 0.58$\pm$0.05 & 5.77$\pm$0.54      & 288$\pm$7       & 249.35$\pm$15.65        & 2.46$\pm$0.52       \\
Briceño-1B                   & 1.97$\pm$0.13     & 16.32$\pm$0.65      & 0.92$\pm$0.04 & 4.74$\pm$0.22      & 430$\pm$9       & 338.50$\pm$21.75        & 2.60$\pm$0.30       \\
Ori-East                     & 2.38$\pm$0.28     & 14.51$\pm$0.69      & 0.78$\pm$0.06 & 8.22$\pm$1.68      & 234$\pm$5       & 127.88$\pm$5.96          & 16.59$\pm$1.17      \\
OBP-Far                      & 6.35$\pm$0.44     & 44.81$\pm$1.37      & 0.85$\pm$0.03 & 14.26$\pm$0.71    & 447$\pm$13     & 430.91$\pm$20.54        & 13.66$\pm$1.72      \\
$\sigma$ Ori                 & 1.12$\pm$0.07     & 12.34$\pm$0.63      & 1.03$\pm$0.03 & 3.04$\pm$0.16      & 554$\pm$31     & 452.98$\pm$33.34        & 3.59$\pm$0.37       \\
OBP-b                        & 5.36$\pm$0.55     & 20.93$\pm$1.13      & 0.59$\pm$0.04 & 8.37$\pm$0.54      & 299$\pm$18     & 319.98$\pm$42.87        & 14.21$\pm$2.23      \\
OBP-d                        & 2.30$\pm$0.16     & 18.44$\pm$0.75      & 0.90$\pm$0.04 & 5.76$\pm$0.28      & 768$\pm$30    & 756.51$\pm$46.34        & 2.44$\pm$0.24       \\
OBP-near A                   & 1.93$\pm$0.35     & 10.39$\pm$0.66      & 0.73$\pm$0.10 & 4.43$\pm$0.72     & 230$\pm$10    & 203.99$\pm$13.16        & 8.96$\pm$2.29       \\
OBP-near B                   & 5.92$\pm$0.52     & 13.55$\pm$0.49      & 0.36$\pm$0.04 & 7.62$\pm$0.43     & 368$\pm$24   & 276.11$\pm$16.49         & 10.50$\pm$2.14      \\
ONC                          & 3.31$\pm$0.13     & 17.72$\pm$0.59      & 0.73$\pm$0.01 & 6.44$\pm$0.21     & 9009$\pm$615  & 7622.87$\pm$654.82     & 1.85$\pm$0.09       \\
Ori-South                    & 5.78$\pm$0.50     & 17.15$\pm$0.70      & 0.47$\pm$0.04 & 7.52$\pm$0.32     & 714$\pm$36   & 420.53$\pm$41.94         & 6.09$\pm$0.48       \\ \hline
\end{tabular}
\end{table*}}

\subsection{Caveats and Limitations}

Our completeness correction assumes that the most populated bin of the observed mass histogram marks the onset of incompleteness and serves as the normalisation point for the PDMF. If this anchor bin is itself incomplete, due to crowding or spatially varying selection in \textit{Gaia} \citep{boubert:2020}, both $N_{\rm s}$ and $M_{\rm cl}$ will still be underestimated. In Section~\ref{sec:results}, we analyse this effect by artificially increasing the mass of the anchor bin by 20\%.

We validated the completeness correction using synthetic clusters in which we imposed incompleteness levels of 25, 50, and 75\% by randomly removing that fraction of stars from the original mass distribution. For each realization, we applied the correction to the resulting incomplete catalog and computed recovery fractions for the number of stars and the stellar mass. Performance was summarized by the median recovery across all realizations at each incompleteness level. The median recovery in the number of stars exceeded \(\sim\)90\% for 25 and 50\% incompleteness and remained \(\gtrsim\)85\% at 75\% incompleteness with a median mass recovery $\gtrsim95$\% in all cases, indicating that the correction is effective. Nevertheless, in highly incomplete cases, especially when the most populated mass bin is affected, both \(N_{\mathrm{s}}\) and \(M_{\mathrm{cl}}\) should be treated as lower limits.

Uncertainties related to the adopted binning configuration and stellar mass estimates may also affect the inferred completeness and cluster masses. We evaluated the robustness of our methodology against these effects through a set of MC tests applied to all clusters in our sample (Appendix~\ref{app:mass_unc}), incorporating variations in the binning configuration and stellar-mass uncertainties. We find that the resulting cluster-mass distributions remain consistent with the adopted values, with median differences typically below 6\% and well within the 1$\sigma$ uncertainties, supporting the robustness of the adopted methodology.

%%%%%
Additionally, we sample only $(\alpha,\delta,\varpi)$ from the KDE when imputing missing stars and deliberately avoid a velocity prior. Because APOGEE and GALAH radial-velocity samples have highly inhomogeneous selection relative to \textit{Gaia} astrometry \citep[e.g.][]{majewski:2017, zasowski:2017, buder:2021}, a velocity prior would reflect survey selection. We therefore preserve the empirical velocity distribution calculated with equation~\eqref{vel_gauss}, preventing biases in $\alpha_{\rm vir}$.
%%%%%

Unresolved stellar multiplicity introduces a potential bias in the derived mass distribution, since unresolved companions can artificially brighten systems and consequently flatten the inferred PDMF, especially at low masses \citep{kroupa:2001,offner:2023}. We mitigate this by applying \textit{Gaia} astrometric-quality filters (low RUWE, high astrometric fidelity) to preferentially retain single-star–like solutions and reject acceleration-affected astrometry \citep{lindegren:2021, rybizki:2022}. These filters reduce, but cannot eliminate, contamination by close or low-contrast binaries, considering that multiplicity is common and mass-dependent in young populations \citep{duchene:2013}, and many binaries remain undetected by \textit{Gaia} over the current baselines \citep{castro-ginard:2024}. Consequently, our corrected memberships and integrated masses should be viewed as conservative lower bounds.

\section{Simulation Ensemble}
\label{sec:sims}

Section~\ref{sec:pdmf} presented the corrected parameters that define the present-day conditions of the clusters in the OSFC. These parameters provide the observational basis for our modelling. However, given the limited three-dimensional constraints on structure and kinematics, we adopt controlled spherical realisations consistent with the global parameters for an OSFC-like scenario. The workflow adopted to build the simulation ensemble involves three main stages: (i) generation of synthetic clusters from dynamical King models, adopting the inferred parameters as initial guesses; (ii) integration of an ensemble of $N$-body realizations; and (iii) analysis of the results. All synthetic clusters and simulations were carried out within the Astrophysical MUltipurpose Software Environment~\citep[\texttt{AMUSE};][]{pelupessy:2013, portegies_book:2018}, a Python-based framework that couples diverse astrophysical solvers for multi-physics experiments.

\subsection{Simulation set-up}

Using the derived structural and dynamical parameters, we built 100 synthetic realisations per cluster by MC sampling the quantities and uncertainties listed in Table~\ref{tab:params_corr}. Initial conditions were generated with the \texttt{AMUSE} King-model sampler, which produces a self-consistent dynamical King distribution function specified by the parameter $W_0$ (derived from the concentration parameter $c$), the virial radius $r_{\rm v}$ and populated with the completeness-corrected membership $N_{\rm cor}$. We note that this approach does not sample the full empirical phase-space PDFs of the observed members; instead, it uses the observed structural constraints to initialise a model that approximates the present-day cluster whose velocities are rescaled to match the measured $\alpha_{\rm vir}$.

For the $i$th realization, we draw $N_i$ stellar masses from the adopted PDMF over the observed mass range. All runs assume single stars, since binary orbital distributions were not derived for this work.

Ensembles were embedded in phase-space using cluster centroids and systemic velocities derived from the catalogue. Observables were transformed to heliocentric Cartesian coordinates with \textsc{Astropy}~\citep{astropy:2022}: positions $(X_h,Y_h,Z_h)$ with $\bmath{Z_h}$ toward the North Galactic Pole, $\bmath{X_h}$ toward the Galactic Centre, and $\bmath{Y_h}$ along Galactic rotation; and velocities $(U_h,V_h,W_h)$ with the same sign convention. Uncertainties were propagated via MC error propagation \citep{Anderson1976} in the observable space and carried through to the centroid and systemic–velocity estimates used to initialise each realisation. We also corrected to the local standard of rest (LSR) by subtracting the solar motion $(U_\odot,V_\odot,W_\odot)=(11.1,12.24,7.25)\ \mathrm{km\,s^{-1}}$ from each stellar velocity \citep{Schoenrich:2010}, and finally transformed to the right-handed Galactocentric frame using a solar distance of $R_0=8.2\ \mathrm{kpc}$~\citep{mcmillan:2017}.

Within \texttt{AMUSE}, we coupled the $N$-body integrator \texttt{Ph4} and the stellar-evolution module \texttt{SeBa} to evolve the clusters self-consistently. \texttt{Ph4} is a fourth-order Hermite scheme with block timesteps, offering high accuracy at reasonable cost for star-cluster dynamics \citep{mcmillan_ph4:2012}, while \texttt{SeBa} evolves the stellar component through parametrized prescriptions \citep{portegies_seba:1996}, which we use to account only for stellar-evolution mass loss and remnant formation by updating particle masses during the integration. Additionally, a Galactic axisimmetric potential was included using the model of \citet{allen:1991}, scaled to a circular speed at the Sun of $v(R_0)=232.8\ \mathrm{km\,s^{-1}}$~\citep{mcmillan:2017}.

Because our simulations include only the stellar component of the OSFC, a key caveat is the presence of residual gas in the youngest systems, which might transiently bias the inferred initial virial state. In this case, we assume that subsequent gas dispersal does not significantly change the currently structural parameters or virial ratios. Therefore, all clusters are evolved from their observed parameters as initial conditions. A full treatment that couples the stellar dynamics to a time-dependent gas potential and tests the sensitivity of $r_{\rm v}$, $c$, and $\alpha_{\rm vir}$ to gas-removal histories is deferred to future work.

Each realisation was evolved for 300 Myr using a fixed time-step of 0.01 Myr, chosen to be well below the shortest cluster crossing time in our sample ($\sim$12 Myr) while keeping the ensemble computationally feasible. Snapshots with full phase-space information were stored every 10 steps. Additionally, we adopted a gravitational softening length of $\sim 0.1$ pc following the prescription in~\citet{portegies_book:2018}; with this setup, the integrations conserve energy to a median fractional error of $10^{-10}$ over the full runs\footnote{All simulations were performed on the Grid UNAM high-performance computing infrastructure at the National Autonomous University of Mexico, using its distributed compute and storage resources, including MinIO platform, widely adopted for workloads in scientific research.}
% in machine learning and scientific research.}.

\subsection{Simulation Analysis}

After completing all realisations, we tracked the phase–space evolution of the OSFC clusters over the full integration. At each snapshot, we determined the centre of mass and systemic velocity of each system. Then we computed for every star the kinetic energy in the cluster–centric frame, $E_K=\tfrac{1}{2}m|{\boldsymbol v}-{\boldsymbol v}_{\rm c}|^2$, and the total gravitational potential energy $E_P$ only of the stellar system. The total energy is $E_T = E_P + E_K$; stars with $E_T<0$ are defined as bound stars in the corresponding snapshot. 
%We applied an iterative boundedness selection: after an initial energy classification, we recomputed the total energy using only the bound subset and repeated until convergence.

Following \citet{dinnbier:2022}, we define a system as a cluster when the bound component contains at least $N_{bound} \geq 10$ stars, adopting this as a practical minimum for defining a gravitationally bound remnant \citep{krumholz:2019}. Additionally, from the bound subset, we derived a set of global diagnostics at each time $t$, principally the number of bound stars $N_{\rm bound}(t)$ and the bound mass $M_{\rm cl}(t)$ which are useful parameters for studying cluster survivability. 

This procedure was applied uniformly to every realisation, and the outputs were collated into time–ordered arrays for all clusters. These datasets underpin the ensemble statistics from which we report medians, 16–84th percentile envelopes, and the most probable dynamical evolutionary paths of the OSFC substructures.

\section{Dynamical Evolution of the OSFC}
\label{sec:results}

\begin{figure*} \centering \includegraphics[width=0.97\textwidth]{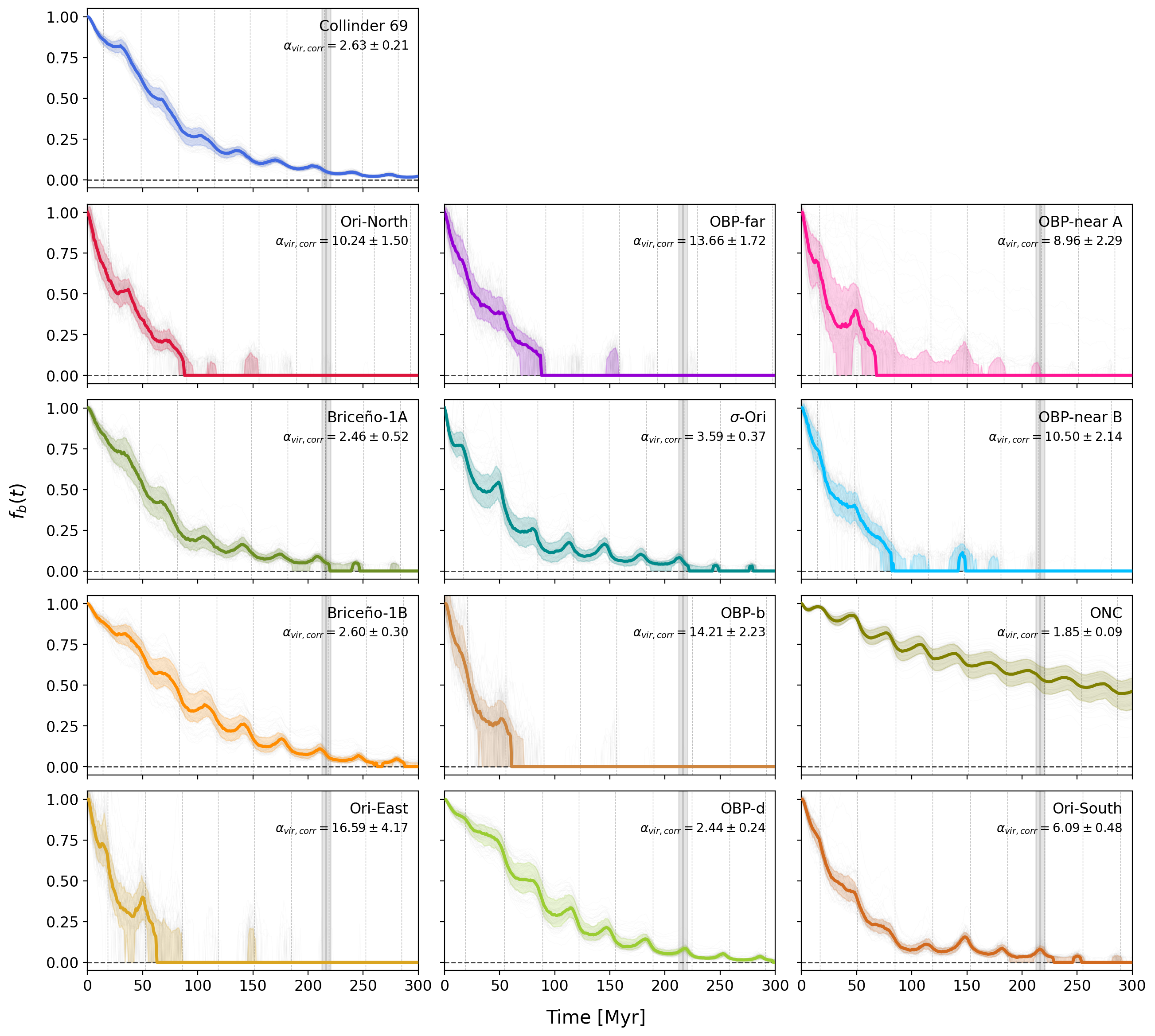}
\caption{Time evolution of the normalised bound fraction, $f_b(t)$ for the 13 OSFC substructures with their respective $\alpha_{vir,cor}$. Each thick coloured curve is the median of 100 direct $N$-body realisations; the shaded band encloses the 16–84th percentiles; and thin grey lines show individual realisations. The horizontal dashed line marks $f_b=0$ and the narrow vertical grey band indicates one Galactic rotation at Orion’s Galactocentric distance, shown as a reference timescale. The vertical dotted lines mark successive mid-plane passages ($z=0$ pc) of the cluster centroid.} 
\label{fig:all_clusters_Ns}
\end{figure*}

\subsection{Evolution of bound stars}
\label{sec:evol_fb}

The previous characterisation in Section~\ref{sec:sims} showed that all OSFC substructures are presently supervirial ($\alpha_{\rm vir,cor}>1$), consistent with the early gas–expulsion–driven expansion observed in \textit{Gaia}-based studies of young clusters~\citep[e.g.][]{kuhn:2019, kounkel_covey:2019, wright:2020, swiggum:2021}. Differences in $\alpha_{\rm vir,cor}$ lead to distinct evolutionary and dissolution timescales, as demonstrated in previous numerical studies~\citep[e.g.][]{lamers:2005, goodwin:2006, baumgardt:2007, kruijssen:2012}. Figure~\ref{fig:all_clusters_Ns} shows the evolution of the normalised bound fraction, $f_b(t)=N_{bound}/N_0$, for all 13 substructures. In each case, we plot the median over all realisations and the 16th–84th percentile envelope. The shaded band indicates one Galactic rotation at Orion’s Galactocentric radius ($227\pm 4$ Myr), computed with a MC error propagation assuming a circular orbit at the OSFC centroid in our adopted potential. The vertical dotted lines mark successive mid-plane passages ($z=0$ pc) of the cluster centroid. The ensemble highlights the diversity of dissolution pathways across Orion in terms of the present-day dynamical state.

The most rapidly dissolving systems show an almost monotonic decay of $f_b$, with complete dissolution well below one Galactic rotation. This subset includes OBP-far, OBP-b, and OBP-North, which are the oldest and most spatially extended clusters in our sample. It also includes OBP-near A and OBP-near B, which belong to a larger ring-like structure reported to be in kinematic expansion \citep{swiggum:2021, sansan:2024}, indicating an already dynamically evolving configuration. These populations, located within the Orion OB1a and OB1b subassociations, may have been shaped by feedback processes associated with winds and supernovae from earlier generations in the region~\citep{bally:2008, kounkel:2018, swiggum:2021}, which could have contributed to the elevated $\sigma_{3D}$ values observed in the hot regime of Figure~\ref{fig:sigma_tau}. Ori-East, although younger, also dissolves on a short timescale; its low mass ($\sim 128\ {\rm M_\odot}$ from Table~\ref{tab:params_corr}) favours an origin as a probable diffuse and weakly bound aggregate~\citep{kruijssen:2012}.

The other subset corresponds to long-lived clusters, with dissolution times comparable to or exceeding one Galactic rotation. The ONC is the clearest example: with the lowest virial parameter ($\alpha_{\rm vir, cor}\approx 1.85$), high stellar mass, and compact structure, it remains bound over 300~Myr, as expected in similar simulations for ONC-like systems~\citep[e.g.][]{kroupa:2001_onc, baumgardtMakino:2003, lamers:2005, kruijssen:2012, Safaei:2025}. Other moderate-supervirial clusters (Collinder~69, Briceño-1A/1B, $\bmath{\sigma}$-Ori, and OBP-d) also seem to retain bound cores, which limit rapid expansion and favour gradual evaporation~\citep[e.g.][]{spitzer:1987, binney:2008}. In the Orion context, these clusters likely represent the densest star-formation events within the complex, having retained sufficient binding energy to survive in a timescale compared to one Galactic rotation \citep{lamers:2005, dinnbier:2022}.

%%%%%
%%%%%

%\subsection{Modulation in the fraction of bound members}
%\label{sec:modulations}

\begin{figure}
    \centering
    \includegraphics[width=0.46\textwidth]{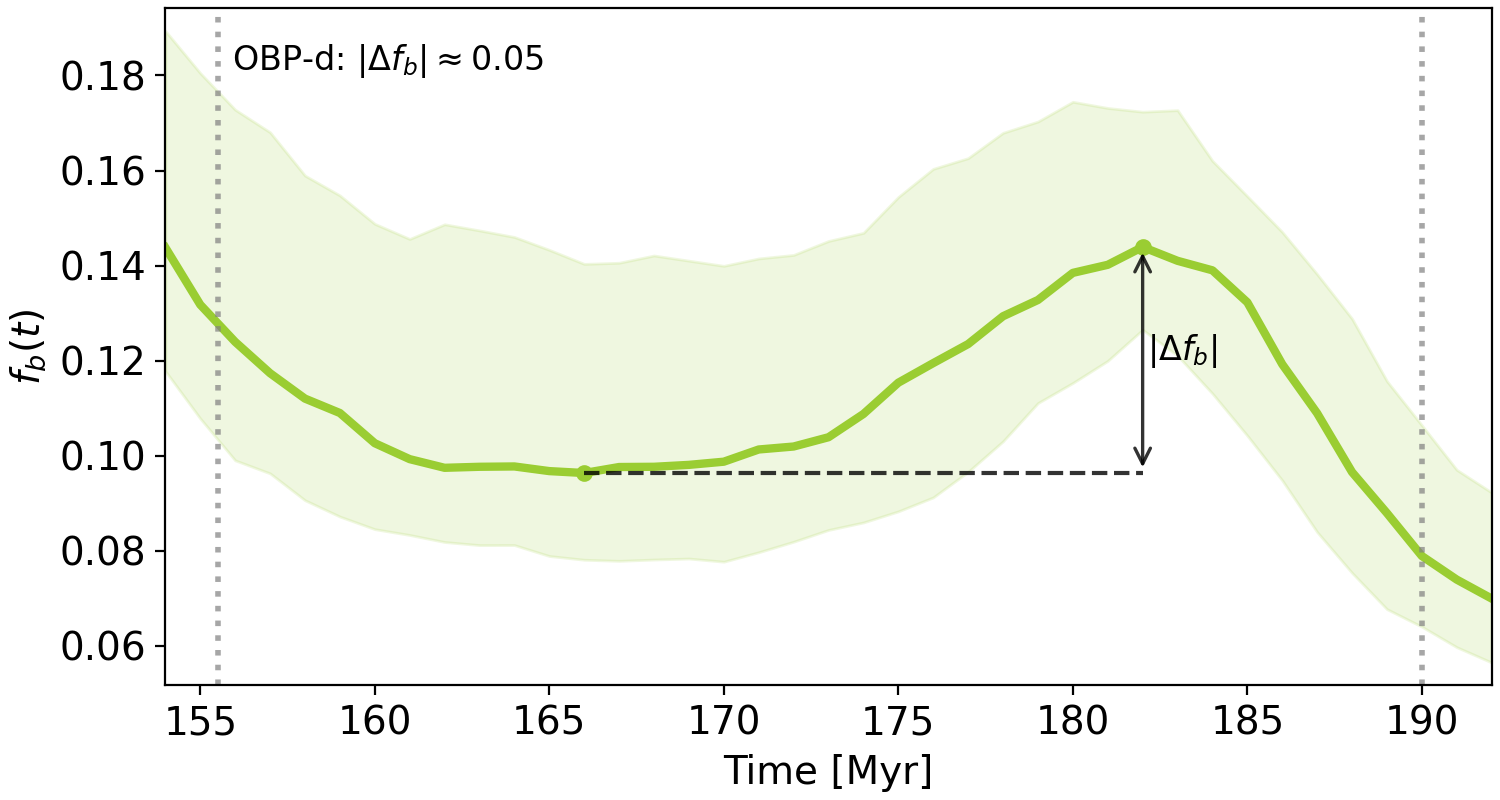}
    \includegraphics[width=0.488\textwidth]{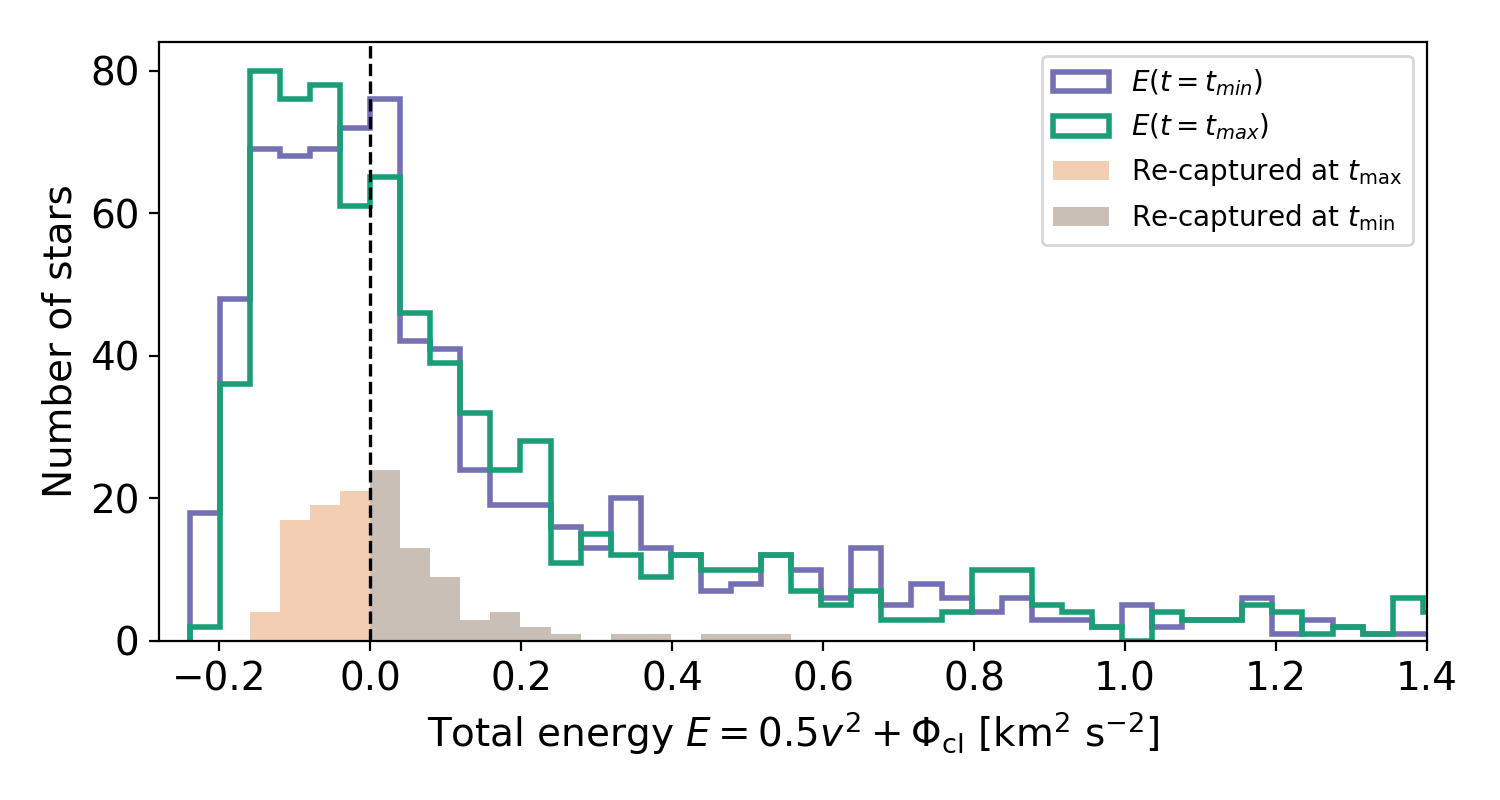}
    \caption{Example of tidal modulation and recapture in OBP--d. \textit{Top panel:} Zoom-in of the bound fraction $f_b(t)$ showing one of the most prominent rebounds. Vertical dotted lines mark Galactic plane passages. \textit{Bottom panel:} Distributions of total energy for the example shown in the top panel at the local minimum, $E(t=t_{\min})$, and subsequent maximum $E(t=t_{\max})$, with shaded histograms highlighting the recaptured subsets. The vertical dashed line marks the boundary $E=0$.}
    \label{fig:recapture}
\end{figure}

All long-lived OSFC clusters display low-amplitude modulations in the bound fraction during dissolution. These ripples are phase-locked to the vertical motion in the Galactic potential with a characteristic timescale of $\sim$ 30--40~Myr. We interpret this as a signature of impulsive disk shocking, where the vertical tidal gradient injects kinetic energy into marginally bound stars, enhancing escape near disk passages, as previously demonstrated in globular cluster studies~\citep[e.g.][]{aguilar:1988, gnedin:1999, Gieles:2008, Kuiper:2010, Webb:2014, Moreno:2024}. In our simulations, the OSFC substructures reach a $|z|_{\max}$ between 84 and 136~pc, implying moderate disk crossings where a fraction of potential escapers linger and can be temporarily reclassified as bound \citep{fukushige:2000, Baumgardt:2001, claydon:2017}. This behaviour suggests that young and moderately supervirial clusters in the OSFC are dynamically sensitive to the vertical component of the Galactic potential, whereas the effect is weak or absent in the highly supervirial systems.

The top panel in Figure~\ref{fig:recapture} illustrates the most prominent modulation for OBP--d, which occurs between 155 and 192 Myr. The median of $f_b(t)$ rises from a local minimum to a subsequent maximum with a recovery fraction of $|\Delta f_{b}|\approx 0.05$. Across all long-lived clusters, simulations suggest an average maximum gain of $|\overline{\Delta f}_{b,max}|\approx 0.04$. The lower panel shows the recapture process in the total energy space, $E_T$. A subset of stars shifts toward more negative $E_T$ after a disk passage, crossing the $E_T=0$ boundary. Since the recaptured population mostly lies close to this threshold, these members are likely to be stripped again at the next disk-crossing shock.

In summary, the OSFC hosts both potentially long- and short-lived stellar substructures along the complex with no apparent spatial segregation. Distinct $f_b(t)$ curves reflect the complex and diverse star-formation history of Orion~\citep[e.g.,][]{bally:2008, zari:2018, chen:2020}. Compact, moderate-virial clusters retain bound cores and show tidal modulations linked to temporary recapture of marginal members, whereas diffuse, supervirial clusters disperse rapidly. These combined behaviours suggest a coherent picture of how the OSFC is dynamically evolving today.
%%%%
%\subsection{Effect of the virial state on system evolution}
\subsection{Dynamical regimes in the OSFC}

The wide spread in dynamical states within the OSFC prompts the question of what mechanism drives current cluster disruption. In this section, we evaluate three possible contributions: cluster–cluster interactions with neighbouring substructures, internal dynamical evolution, and the influence of the Galactic tidal field.

To test the first contribution, we ran a control experiment in which each cluster was individually evolved within the axisymmetric Milky Way potential, and compared the resulting $f_b(t)$ with those in Figure~\ref{fig:all_clusters_Ns}. We found that in all cases, the bound-fraction evolution and centroid trajectories are statistically equivalent, indicating weak interaction. The obtained median $f_b(t)$ curves are nearly identical with an average root-median-squared error $\lesssim  0.03$, well below the intrinsic dispersion of the MC percentiles. This is consistent with an impulsive regime in which the relative cluster speeds exceed their internal velocity dispersions \citep{delafuente:2009, gavagnin:2016, pang:2021_2}.

\begin{figure}
    \centering
    \includegraphics[width=0.493\textwidth]{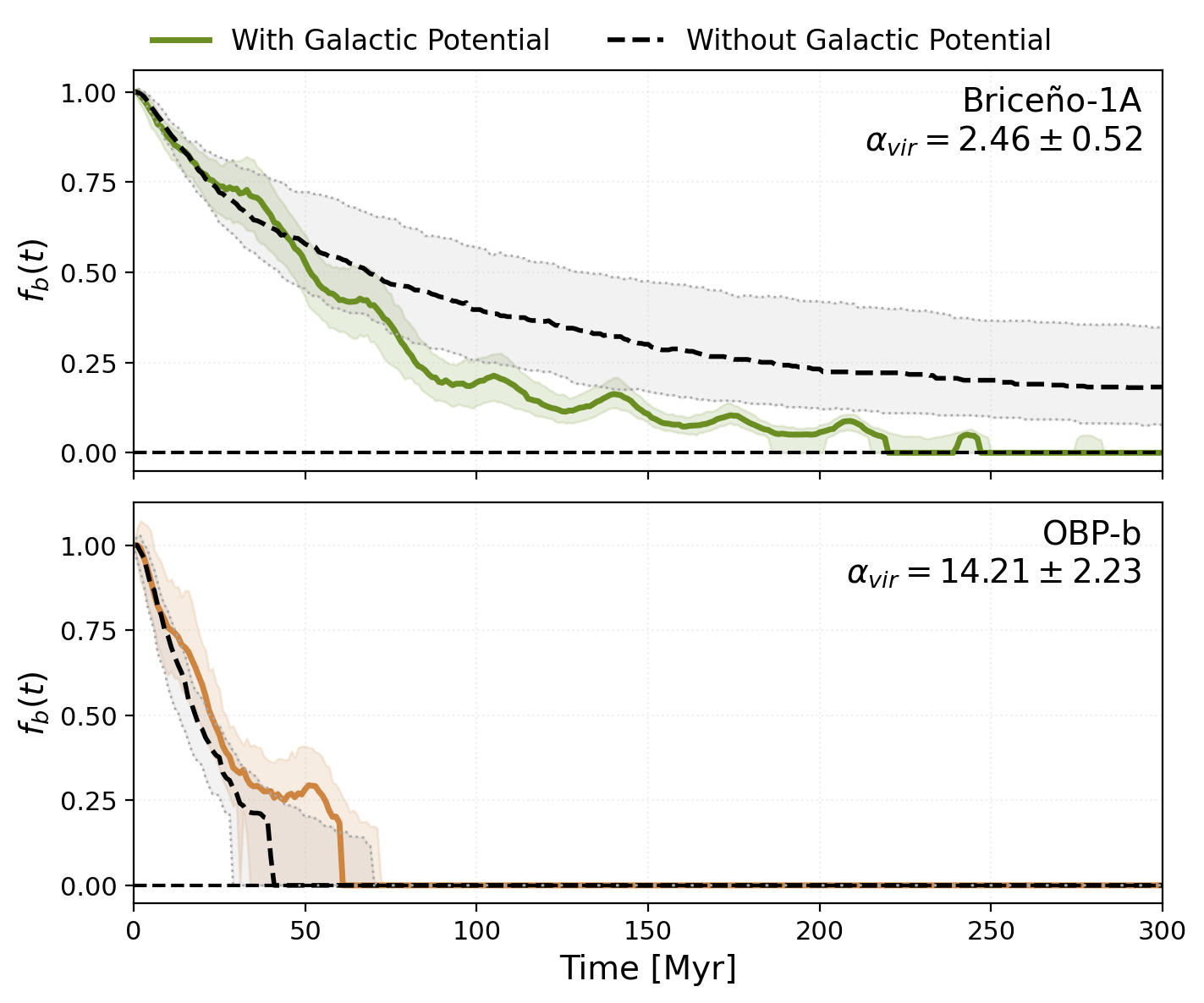}
    \caption{Evolution of the bound fraction $\bmath{f_b(t)}$ for two representative OSFC groups: Briceño–1A (top) and OBP–b (bottom). Coloured solid lines show the median evolution including the Galactic potential. Black dashed lines correspond to runs without the Galactic potential, and light grey curves show individual realisations for reference. In both cases, shaded regions indicate the 16–84th percentiles across realisations}
    \label{fig:comp_disolution}
\end{figure}

For the remaining two contributions, we performed an additional ensemble without the Galactic potential, isolating the evolution driven by the stellar–evolution mass loss, two–body relaxation, and supervirial expansion over $300~\mathrm{Myr}$. Figure~\ref{fig:comp_disolution} compares two representative cases (Briceño-1A and OBP-b), spanning moderate and high values of $\alpha_{\rm vir,cor}$, which highlights the transition from a Galactic-potential–regulated to an internal-dynamics–dominated regime. 

In moderate-supervirial clusters (e.g. Briceño-1A), $f_b(t)$ declines systematically faster when the Galactic potential is included, as expected for clusters evolving in an external tidal field \citep[e.g.][]{baumgardtMakino:2003}. The divergence between tidal and non-Galactic runs becomes evident after $\sim$20 Myr; at this stage, the vertical modulation of $f_b(t)$ becomes pronounced, consistent with the cluster having crossed the Galactic plane for the first time and subsequently reaching ${|z|_{\rm max}}$, where partial recapture of marginally unbound stars can occur. The sustained offset thereafter indicates that the external field regulates the long-term evolution. In contrast, highly supervirial systems (e.g. OBP-b) exhibit a rapid and nearly identical decline of $f_b(t)$ with and without the Galactic potential, reaching dissolution on comparable timescales. Although brief recapture episodes can slightly delay dissolution in the Galactic-potential run, the overall evolution remains dominated by internal expansion, indicating that excess internal kinetic energy drives the rapid unbinding \citep{hills:1980, goodwin:2006}.

Taken together, these comparisons suggest an OSFC-specific classification of dissolution modes. Moderate-supervirial clusters occupy a Galactic-potential-regulated regime, whereas highly supervirial systems are internal-dynamics--dominated. Cluster--cluster interactions are negligible within the explored parameter space, as the isolated and full-complex runs are statistically indistinguishable. This framework captures the present dynamical state of the OSFC under its current Galactic conditions.

%%%%SECTION 6.3. STARTS HERE
%\subsection{$T_{\rm dis}$-- $\alpha_{\rm vir}$ diagnostic plane}
%\label{sec:Tdis_alpha}

\begin{figure*}
    \centering
    \includegraphics[width=0.97\textwidth]{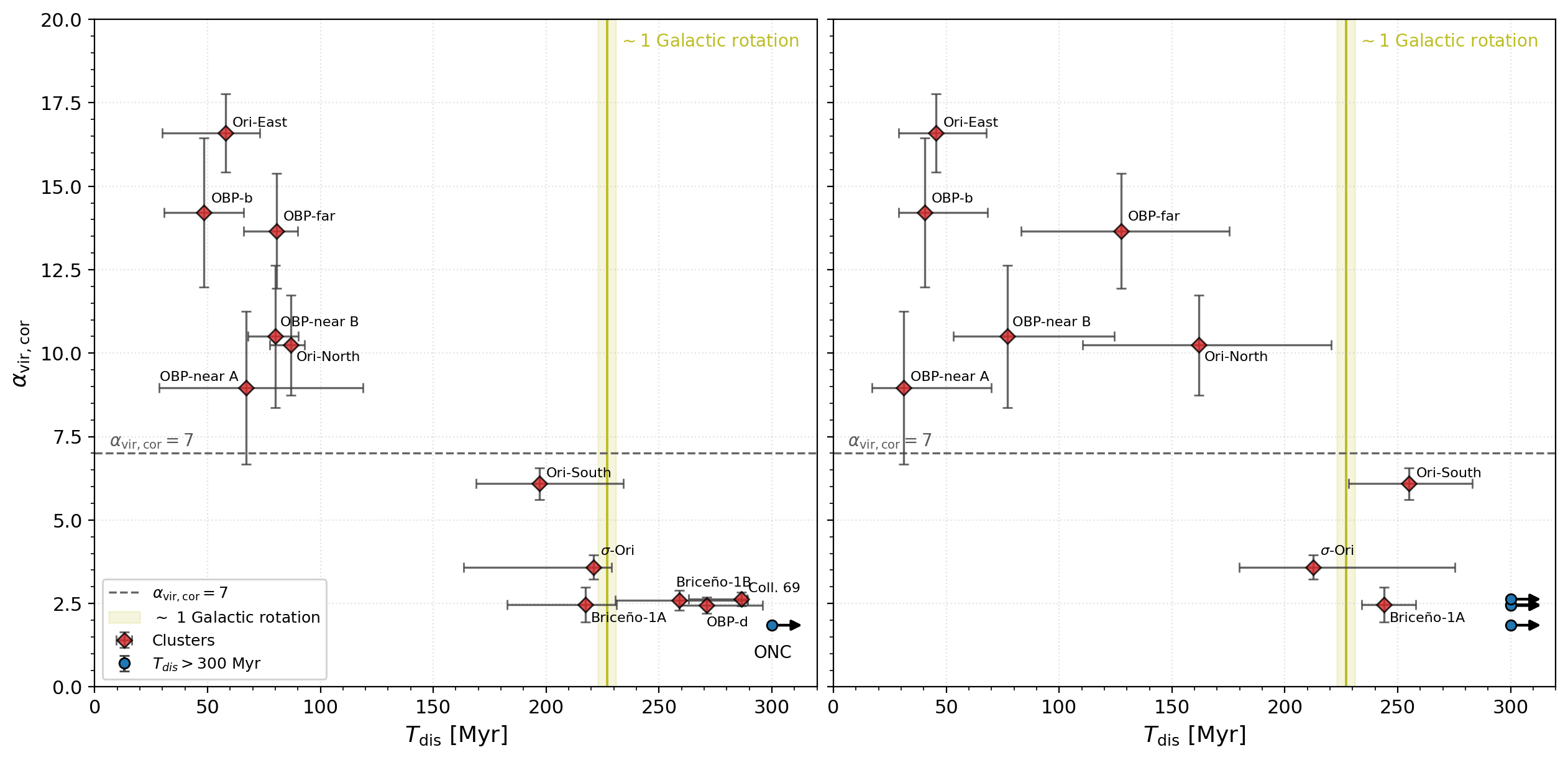}
    \caption{Dissolution time $T_{\rm dis}$ versus present–day virial ratio $\alpha_{\rm vir}$ for Orion substructures. Red diamonds show the median $T_{\rm dis}$ over the ensemble; horizontal error bars denote the 16th--84th percentiles in $T_{\rm dis}$ and vertical bars the $1\sigma$ uncertainty in $\alpha_{\rm vir}$. The dashed horizontal line marks $\alpha_{\rm vir}=7$ just for reference. The shaded vertical band (with central line) indicates 1 Galactic rotation at Orion’s position. \textit{Left-panel}: Runs with the Galactic potential and \textit{Right-panel} without the Galactic potential.}
    \label{fig:dis_alpha}
\end{figure*}

Figure~\ref{fig:dis_alpha} provides a diagnostic of the OSFC outcomes by relating the dissolution time, $T_{\rm dis}$ (defined as the first epoch when the median bound curve satisfies $N_{\rm bound}<10$), with the present-day $\alpha_{\rm vir,cor}$. In the runs including the Galactic potential (left panel), we distinguish both dynamical regimes: clusters with $\alpha_{\rm vir,cor}\gtrsim 7$ might have dissolution timescales between $T_{\rm dis}\sim$30--120~Myr, whereas clusters with $\alpha_{\rm vir,cor}\lesssim 7$ typically retain bound cores beyond $\sim$160~Myr. Notably, the region around $\alpha_{\rm vir,cor}\approx 7$ and $T_{\rm dis}\approx 150$~Myr leaves a visible gap in the OSFC sample. We suggest that $\alpha_{\rm vir}\approx 7$ should represent the transition between the internal-dynamics--dominated and Galactic-potential-regulated regime under Orion conditions. Clusters near this boundary might be sensitive to small variations in $\bmath{\alpha_{\rm vir,cor}}$, as well as to secondary parameters such as mass or concentration. This suggests that small differences in these properties might shift a system from short- to long-lived clusters.

For comparison, the right panel of Figure~\ref{fig:dis_alpha} shows the run without the Galactic potential. As expected, the dissolution times of several substructures are extended, particularly the low–$\alpha_{\rm vir}$ clusters (e.g. Collinder~69, Briceño-1B, OBP-d, and the ONC), which remain bound beyond 300 Myr. Other groups, such as Ori-South and Briceño-1A, show an increase in their $T_{dis}$ by several tens of Myr. Finally, massive and highly supervirial systems Ori-North and OBP-far ($\sim 398~{\rm M}_\odot$ and $\sim 430~{\rm M}_\odot$ from Table 
~\ref{tab:params_corr}, respectively) exhibit comparatively longer survival, with $T_{dis}$ between $\sim$ 80 and 220 Myr, suggesting that the Galactic potential slightly accelerates their disruption, whereas lower-mass supervirial groups are unaffected, as internal expansion dominates.

Overall, the $\alpha_{\rm vir}-T_{dis}$ plane provides an empirical diagnostic of cluster evolution in Orion. Recent works have discussed the limited exploration of virial analyses in observed young clusters and the development of simulations exploring highly expanding systems~\citep{wright:2024, dellacroce:2024}, whereas classical studies generally infer $T_{dis}$ from cluster demographics~\citep[e.g.][]{baumgardtMakino:2003, lamers:2005, goodwin:2006, kruijssen:2012}. In this context, our results combine observational and theoretical perspectives within a uniform framework to predict the fate of OSFC clusters.

\subsection{Analysis of incompleteness in the highest mass bin}
\label{sec:twenty_percent_test}

\begin{figure}
\centering
\includegraphics[width=0.46\textwidth]{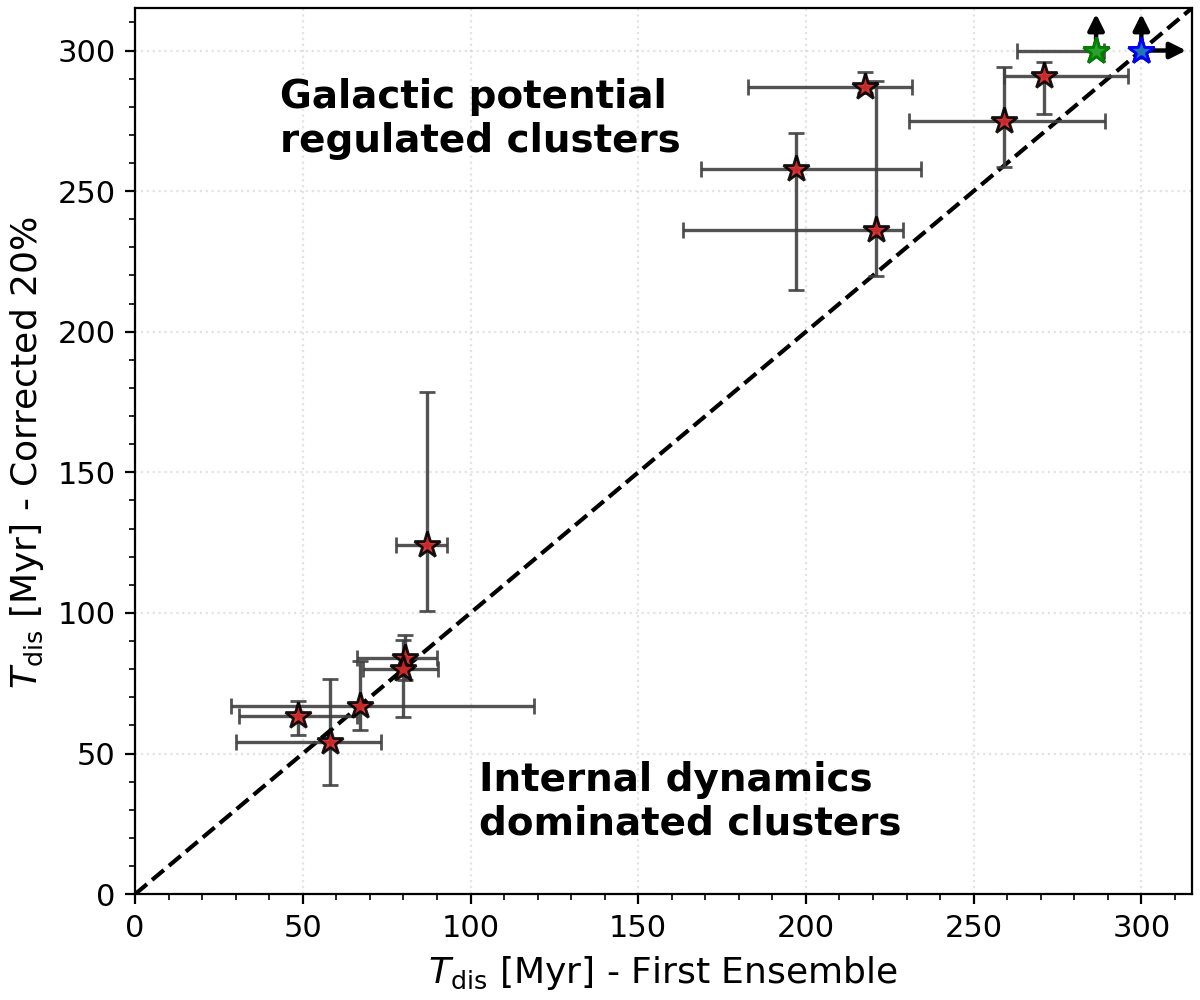}
\caption{Comparison of dissolution times with and without an extra correction in the highest bin by an additional $20\%$. The dashed line is the identity relation and error bars the 16--84th percentiles from the realizations. The blue star (ONC) shows a $T_{\rm dis}>300~\mathrm{Myr}$. The green star (Collinder~69) crosses the censoring threshold and dissolves after $300~\mathrm{Myr}$ in the $+20\%$ case.}
\label{fig:comp_dissolution_50p}
\end{figure}

In Section~\ref{sec:pdmf}, we corrected the PDMFs of the Orion substructures by fitting a canonical IMF and resampling to account for magnitude-limited incompleteness (e.g. \citealt{kroupa:2001,chabrier:2003}).  Although normalizing the PDMFs to the highest-mass bin helps minimize the loss of stars selected as members, a  limitation of this approach is that this bin may itself be incomplete due to selection effects (saturation, crowding, unresolved multiplicity). To assess the impact of this possible effect, we repeated the construction of initial conditions assuming that the top mass bin is still incomplete by $20\%$, and re-ran the full $N$-body ensemble. The assumed incompleteness was chosen arbitrarily to explore the general effect of selection biases on our method and the predictions about the survival of the clusters. In Figure~\ref{fig:comp_dissolution_50p}, we compare the resulting $T_{\rm dis}$ to the baseline on a one–to–one plot.

Two particular behaviours emerge. First, internal-dynamics–dominated clusters lie very close to the identity line: their $T_{\rm dis}$ is essentially unchanged by the $+20\%$ correction. This confirms that their fate is set by the initial virial imbalance and rapid internal expansion on a few Myrs, with small adjustments to the high–mass tail having negligible impact. We have, as the only exception, the group Ori-North, which is expected to dissolve between 100 and 180 Myr. Second, Galactic-potential–regulated clusters (moderate $\alpha_{\rm vir,cor}$) shift systematically above the line. Therefore, median lifetimes increase by several tens of Myr, with a maximum gain of $\sim60~\mathrm{Myr}$. Here, the extra initial binding slightly deepens the potential, delaying stripping by the Galactic tide; the effect is modest because early stellar–evolution mass loss erodes part of the added binding within $\sim 3$–$40~\mathrm{Myr}$.

Two long-lived systems show instructive exceptions. The ONC (blue star in Figure~\ref{fig:comp_dissolution_50p}) remains right–censored ($T_{\rm dis}>300~\mathrm{Myr}$) in both ensembles. In contrast, Collinder~69 (green star) crosses the censoring threshold in the $+20\%$ ensemble and remains bound within our 300~Myr integration window, consistent with the systematic lifetime increase seen for the other Galactic-potential--regulated clusters when their initial mass is enhanced.

Overall, the dynamical regime of each group remains unchanged: high–$\alpha_{\rm vir}$ systems remain internal-dynamics–dominated with nearly identical $T_{\rm dis}$, whereas moderate–$\alpha_{\rm vir,cor}$ clusters retain longer lifetimes with modest sensitivity ($\lesssim 60~\mathrm{Myr}$).

\subsection{Limitations}
Our modelling omits several physical ingredients. 
(i) We did not include primordial binaries, mainly due to computational constraints. 
Binary heating (hard–soft interactions and three-body encounters) injects internal energy and can accelerate expansion, potentially shortening $T_{\rm dis}$ for otherwise long–lived clusters \citep[e.g.][]{goodwin:2006}. The magnitude of the effect depends on the initial binary fraction and orbital parameter distributions~\citep{kroupa:1995}. Without tailored binary populations, we cannot quantify it here, so the reported bound fractions and lifetimes should be regarded as upper limits for systems with significant primordial binarity~\citep{heggie:2006}. 
(ii) For very young systems such as the ONC, we neglected any residual gas potential that may bias the initial binding energy low (making the stellar component appears more supervirial), while time-dependent gas expulsion can drive additional early expansion. This may primarily affect the first $\sim$1-5 Myr~\citep{dinnbier-walch:2020}.
(iii) We adopted a steady, axisymmetric Galactic potential \citep{allen:1991}, omitting bar, spiral structure, and other features. This likely underestimates the variance in tidal heating and could make our lifetimes upper limits for clusters on orbits with strong perturbations \citep[e.g.][]{gieles:2006}. (iv) Due to observational limitations, we adopted spherical King initial conditions matched to global observables, since the OSFC clusters might show complex and non-spherical configurations with incomplete 6D phase-space; the simulations should therefore be viewed as controlled, global-evolution experiments.

\section{Conclusions}
\label{sec:conclusions}

We combined \textit{Gaia}-based information, radial velocity data from spectroscopic surveys (APOGEE and GALAH), completeness–corrected PDMFs, and an ensemble of $N$-body simulations in a steady axisymmetric Galactic potential to trace the evolution of young stellar systems across the OSFC. Our main conclusions are:

\begin{itemize}
\item From the observational identification performed by \citet{sansan:2024}, we found that all \textit{Big Structure} clusters in the OSFC are in a supervirial state, with $\alpha_{\rm vir}>1$. The spread in $\alpha_{\rm vir}$ indicates a wide range of current dynamical states within the same complex, from mildly supervirial to extremely supervirial clusters.

\item We corrected the PDMFs for selection/quality cuts and other observational limits. From these corrections, we estimated that the observed samples have $\sim40\%$ of the likely total stellar mass on average. Consequently, the observed $\alpha_{\rm vir}$ might be overestimated by $\simeq 34\%$ on average; however, even after corrections, all systems remain supervirial. Thus, the conclusion that the Orion substructures are out of virial equilibrium is robust against sample incompleteness.

\item Our results suggest that OSFC clusters can be separated into two dynamical regimes: Long-lived clusters (ONC, Collinder~69, Briceño-1A/B, OBP-d and $\bmath{\sigma}$-Ori) occupy a Galactic-potential–regulated regime, where the Galactic tide modulates the dissolution. In contrast, short-lived clusters (Ori-North, OBP-far, OBP-b, OBP-near A/B, and Ori-East) evolve in an internal-dynamics–dominated regime, where internal kinematics drives dissolution. 

\item Cluster–cluster interactions are negligible within our explored parameter space. Evolving each cluster individually or in the presence of neighbouring substructures produces statistically indistinguishable bound fraction evolution within our MC uncertainties. This behaviour is consistent with an impulsive-encounter regime in which relative inter-cluster velocities exceed the internal velocity dispersions of the clusters.

\item Long-lived clusters show low-amplitude, phase-locked modulations in $f_b(t)$ correlated with $z(t)$, consistent with disk-crossing tidal heating and temporary recapture of marginal members. These $z$–linked modulations are the hallmark of the Galactic-potential-regulated regime. This suggests that young and moderately bound systems in the OSFC are dynamically sensitive to the vertical structure of the Galactic potential.

\item Our results suggest a separation in the $T_{\rm dis}$–$\alpha_{\rm vir}$ plane: OSFC clusters with $\alpha_{\rm vir,cor}\lesssim7$ typically have a $T_{\rm dis}\gtrsim160$~Myr, whereas clusters with $\alpha_{\rm vir,cor}\gtrsim7$ have an expected dissolution around $T_{\rm dis}\sim$ 30--120~Myr. This behaviour is consistent with previous numerical studies showing that the current virial state is a key determinant of cluster survival. The apparent separation around 150 Myr suggests a transition point at $\alpha_{\rm vir,cor}\approx 7$ between long- and short-lived clusters.

\item Stress tests imposing a 20\% excess in the highest-mass bin lead to minor quantitative changes in the overall dynamical behaviour. Clusters in the moderate-virial regime dissolve on comparable timescales, while highly supervirial systems remain dominated by rapid expansion. We therefore conclude that our main results and inferred dissolution timescales are robust against residual PDMF incompleteness at this level.

\item The coexistence of long- and short-lived systems within a single complex reflects Orion’s heterogeneous history. The OSFC spans a broad range of ages, kinematics, and substructure, consistent with multiple epochs of star formation shaped by feedback (e.g. winds and supernovae), which can remove gas and drive strongly supervirial expansion in some groups while leaving others with a surviving bound core.
\end{itemize}

The framework applied in this work can be expanded to evaluate the dynamical evolution of other nearby complexes. For future work, we want to incorporate primordial binaries, quantify RV selection effects, include other perturbers (giant molecular clouds and spiral–arm passages) along measured orbits, and test for initial rotation or anisotropy. These refinements would provide a more accurate estimate of the expected $T_{dis}$.

Young clusters are key dynamical laboratories for Galactic evolution because they trace the transition from compact clusters to the Galactic stellar field via gas clearing, expansion, relaxation, and tidal stripping. 
Combining present-day phase-space with direct $N$-body modelling yields predictive timelines for survival and dispersal into the field. In this process, two practical obstacles remain: (i) sample completeness in the recovery of faint, reliable members and extended outskirts for robust structural characterisation, and (ii) complete and precise radial velocities for a better estimate of $\alpha_{\rm vir}$. Forthcoming catalogues such as \textit{Gaia} DR4 and the Legacy Survey of Space and Time (LSST) will help mitigate both limitations, enabling higher-fidelity structure fits, better-initialised ensembles, and improved estimates of dissolution timescales, turning star-forming complexes into benchmarks for predictive cluster dynamics.

\section*{Acknowledgements}

We thank the anonymous referee for the careful review, which helped us improve this work. S.S.-S. is supported by CONAHCyT Beca Nacional de Posgrado. S.S.-S. and A.P.-V. acknowledge the DGAPA–PAPIIT grant  IA103224 and IN112526. J. H. and L.A. acknowledge support from the DGAPA–PAPIIT grant, IG-101723 and IN110126.
The authors acknowledge CONAHCyT grant 86372 entitled ‘Citlalcóatl: a multiscale study at the new frontier of the formation and early evolution of stars and planetary systems', Mexico.

This work has made use of data from the European Space Agency (ESA) mission
{\it Gaia} (\url{https://www.cosmos.esa.int/gaia}), processed by the {\it Gaia}
Data Processing and Analysis Consortium (DPAC,
\url{https://www.cosmos.esa.int/web/gaia/dpac/consortium}). Funding for the DPAC
has been provided by national institutions, in particular the institutions
participating in the {\it Gaia} Multilateral Agreement.

This research was performed using services/resources provided by Grid UNAM, which is a collaborative effort driven by DGTIC and the research institutes of Astronomy, Nuclear Science and Atmosphere Science and Climate Change at UNAM.
%%%%%%%%%%%%%%%%%%%%%%%%%%%%%%%%%%%%%%%%%%%%%%%%%%
\section*{Data Availability}
 
The \textit{Gaia} DR3, APOGEE-2 and GALAH-DR3 data sets used for this work are publicly available. The final data sets for the clusters recovered and simulations will be shared upon reasonable request to the corresponding author.

%%%%%%%%%%%%%%%%%%%% REFERENCES %%%%%%%%%%%%%%%%%%

% The best way to enter references is to use BibTeX:

\bibliographystyle{mnras}
\bibliography{references} % if your bibtex file is called example.bib

% Alternatively you could enter them by hand, like this:
% This method is tedious and prone to error if you have lots of references
%\begin{thebibliography}{99}
%\bibitem[\protect\citeauthoryear{Author}{2012}]{Author2012}
%Author A.~N., 2013, Journal of Improbable Astronomy, 1, 1
%\bibitem[\protect\citeauthoryear{Others}{2013}]{Others2013}
%Others S., 2012, Journal of Interesting Stuff, 17, 198
%\end{thebibliography}

%%%%%%%%%%%%%%%%%%%%%%%%%%%%%%%%%%%%%%%%%%%%%%%%%%

%%%%%%%%%%%%%%%%% APPENDICES %%%%%%%%%%%%%%%%%%%%%

\appendix

\section{Radial Density Profile Fittings}
\label{app:rdps_params}

Here, we present in Table~\ref{tab:rdp_fitting} the results of fitting RDPs of Orion clusters using two widely applied models: the King profile and the Elson–Fall–Freeman (EFF) profile. The King profile provides estimates of the central surface density, core radius, and truncation radius, while the EFF profile captures the extended structure of young, non-truncated clusters through its characteristic scale length and slope parameter $\gamma$. By fitting both models in parallel, we are able to compare the structural descriptions of each cluster and quantify their central concentrations, outer extents, and overall distributions. %These results form the structural foundation for subsequent dynamical analysis, including virial radius estimates and survival predictions.

{\renewcommand{\arraystretch}{1.3}
\begin{table*}
\centering
\caption{Best-fitting parameters of the radial density profiles for Orion subclusters. Columns 2–4 list the central surface density ($\Sigma_{o,k}$), core radius ($r_c$), and truncation radius ($r_t$) derived from King profile fits. Columns 5–7 provide the central surface density ($\Sigma_{o,e}$), scale parameter ($a$), and slope parameter ($\gamma$) obtained from EFF profile fits. Reported uncertainties correspond to the 16th–84th percentile ranges from the posterior distributions of the fits.}
\label{tab:rdp_fitting}
\begin{tabular}{lcccccc}
\hline
\multicolumn{1}{c}{Cluster}  & \begin{tabular}[c]{@{}c@{}}$\Sigma_{o,k}$\\ (stars deg$^{-2}$) $\times 10^2$\end{tabular} & \begin{tabular}[c]{@{}c@{}}$r_c$\\ (deg)\end{tabular} & \begin{tabular}[c]{@{}c@{}}$r_t$\\ (deg)\end{tabular} & \begin{tabular}[c]{@{}c@{}}$\Sigma_{o,e}$\\ (stars deg$^{-2}$) $\times 10^2$\end{tabular} & \begin{tabular}[c]{@{}c@{}}$a$\\ (deg)\end{tabular} & $\gamma$               \\
\multicolumn{1}{c}{(1)}      & (2)    & (3)     & (4)     & (5)    & (6)    & (7)   \\ \hline \hline
$\lambda$ Ori (Collinder 69) & 3.70$^{+0.38}_{-0.30}$   & 0.72$^{+0.25}_{-0.16}$   & 2.45$^{+0.09}_{-0.09}$    & 3.80$^{+0.50}_{-0.52}$    & 0.44$^{+0.08}_{-0.06}$    & 3.24$^{+0.34}_{-0.27}$ \\
Ori-North                               & 1.05$^{+0.08}_{-0.07}$    & 1.24$^{+0.17}_{-0.14}$   & 3.88$^{+0.09}_{-0.09}$    & 0.68$^{+0.10}_{-0.09}$    & 1.13$^{+0.27}_{-0.19}$    & 3.67$^{+0.76}_{-0.53}$ \\
Briceño-1A                            & 2.18$^{+0.22}_{-0.22}$    & 0.54$^{+0.08}_{-0.06}$   & 2.02$^{+0.08}_{-0.09}$    & 1.49$^{+0.22}_{-0.22}$    & 0.69$^{+0.20}_{-0.15}$    & 4.73$^{+1.68}_{-1.00}$ \\
Briceño-1B                            & 4.37$^{+0.38}_{-0.39}$    & 0.28$^{+0.02}_{-0.02}$   & 2.51$^{+0.10}_{-0.09}$    & 3.96$^{+0.40}_{-0.39}$    & 0.41$^{+0.06}_{-0.05}$    & 3.85$^{+0.50}_{-0.37}$ \\
Ori-East                                 & 3.94$^{+0.52}_{-0.50}$    & 0.22$^{+0.03}_{-0.02}$    & 1.81$^{+0.10}_{-0.10}$    & 3.74$^{+0.53}_{-0.52}$    & 0.20$^{+0.04}_{-0.04}$    & 2.59$^{+0.39}_{-0.30}$ \\
OBP-Far                                & 0.97$^{+0.10}_{-0.09}$    & 0.53$^{+0.04}_{-0.04}$    & 5.18$^{+0.10}_{-0.11}$    & 0.84$^{+0.09}_{-0.09}$    & 0.88$^{+0.21}_{-0.14}$    & 4.11$^{+0.93}_{-0.57}$ \\
$\sigma$ Ori                          & 9.15$^{+0.92}_{-0.95}$    & 0.16$^{+0.01}_{-0.01}$    & 1.75$^{+0.10}_{-0.10}$    & 8.41$^{+0.91}_{-0.85}$    & 0.23$^{+0.04}_{-0.03}$    & 3.64$^{+0.59}_{-0.40}$ \\
OBP-b                                   & 1.54$^{+0.13}_{-0.13}$     & 0.74$^{+0.08}_{-0.07}$    & 2.55$^{+0.09}_{-0.09}$    & 1.07$^{+0.10}_{-0.10}$    & 1.44$^{+0.12}_{-0.17}$    & 8.79$^{+0.89}_{-1.39}$ \\
OBP-d                                   & 4.86$^{+0.45}_{-0.42}$     & 0.37$^{+0.04}_{-0.03}$    & 2.26$^{+0.09}_{-0.09}$    & 4.06$^{+0.47}_{-0.44}$    & 0.46$^{+0.07}_{-0.06}$    & 3.76$^{+0.46}_{-0.35}$ \\
OBP-Near A                          & 2.62$^{+0.36}_{-0.32}$     & 0.39$^{+0.08}_{-0.06}$    & 1.70$^{+0.09}_{-0.10}$    & 1.93$^{+0.69}_{-0.40}$    & 0.42$^{+0.21}_{-0.13}$    & 3.69$^{+1.66}_{-0.82}$ \\
OBP-Near B                          & 1.42$^{+0.15}_{-0.15}$     & 0.85$^{+0.11}_{-0.10}$    & 1.99$^{+0.07}_{-0.07}$    & 0.87$^{+0.08}_{-0.08}$    & 1.44$^{+0.13}_{-0.19}$    & 8.64$^{+0.96}_{-1.53}$ \\
ONC                                      & 21.10$^{+0.79}_{-0.75}$   & 0.50$^{+0.03}_{-0.02}$    & 2.58$^{+0.06}_{-0.06}$    & 18.60$^{+0.70}_{-0.72}$   & 1.04$^{+0.10}_{-0.08}$   & 6.87$^{+0.73}_{-0.61}$ \\
Ori-South                               & 3.33$^{+0.25}_{-0.23}$    & 0.87$^{+0.12}_{-0.10}$     & 2.27$^{+0.06}_{-0.07}$    & 3.33$^{+0.25}_{-0.23}$     & 2.82$^{+0.21}_{-0.17}$    & 7.72$^{+1.53}_{-1.58}$ \\ \hline
\end{tabular}
\end{table*}
}

\section{Velocity Distribution Fittings}
\label{app:velocity}
In this appendix we present the Gaussian fits to the velocity distributions of the Orion clusters seen in Table~\ref{tab:velocity_fit}. For each system, we modelled the radial velocity ($v_{RV}$) and tangential velocity components ($v_{\alpha}$, $v_{\delta}$) with a trivariate Gaussian distribution, obtaining both the median velocities and their dispersions. This approach provides a consistent statistical characterization of the internal kinematics of each group, accounting for observational uncertainties in both radial velocity and proper motion. From these fits, we derived the total three-dimensional velocity dispersion $\sigma_T$, which serves as a key diagnostic of the clusters’ dynamical states and was later used to constrain their virial ratios and survival timescales.

{\renewcommand{\arraystretch}{1.3}
\begin{table*}
\centering
\caption{Best-fitting velocity parameters for Orion subclusters. Columns 2–4 give the median velocities in the radial ($v_{RV}$), right ascension ($v_{\alpha}$), and declination ($v_{\delta}$) components. Columns 5–7 report the corresponding velocity dispersions ($\sigma_{RV}$, $\sigma_{\alpha}$, $\sigma_{\delta}$) derived from the Gaussian fits. Column 8 lists the total three-dimensional velocity dispersion $\sigma_T$. Reported uncertainties correspond to the 16th–84th percentile ranges of the posterior distributions.}
\label{tab:velocity_fit}
\begin{tabular}{lccccccc}
\hline
\multicolumn{1}{c}{Cluster} & \begin{tabular}[c]{@{}c@{}}$v_{RV}$\\ (km s$^{-1}$)\end{tabular} & \begin{tabular}[c]{@{}c@{}}$v_{\alpha}$\\ (km s$^{-1}$)\end{tabular} & \begin{tabular}[c]{@{}c@{}}$v_{\delta}$\\ (km s$^{-1}$)\end{tabular} & \begin{tabular}[c]{@{}c@{}}$\sigma_{RV}$\\ (km s$^{-1}$)\end{tabular} & \begin{tabular}[c]{@{}c@{}}$\sigma_{\alpha}$\\ (km s$^{-1}$)\end{tabular} & \begin{tabular}[c]{@{}c@{}}$\sigma_{\delta}$\\ (km s$^{-1}$)\end{tabular} & \begin{tabular}[c]{@{}c@{}}$\sigma_{T}$\\ (km s$^{-1}$)\end{tabular} \\
\multicolumn{1}{c}{(1)}     & (2)     & (3)    & (4)    & (5)   & (6)   & (7)    & (8)    \\ \hline \hline
$\lambda$ Ori (Collinder69)    & 27.11$^{+0.14}_{-0.12}$    & 2.13$^{+0.06}_{-0.06}$   & $-$3.95$^{+0.03}_{-0.03}$    & 0.86$^{+0.13}_{-0.11}$   & 0.94$^{+0.05}_{-0.05}$   & 0.46$^{+0.02}_{-0.02}$   & 1.36$^{+0.11}_{-0.10}$   \\
Ori-North        & 31.39$^{+0.33}_{-0.32}$     & $-$1.17$^{+0.04}_{-0.04}$    & 1.31$^{+0.03}_{-0.03}$   & 2.39$^{+0.32}_{-0.32}$   & 0.64$^{+0.03}_{-0.03}$    & 0.46$^{+0.03}_{-0.02}$   & 2.51$^{+0.30}_{-0.29}$     \\
Briceño-1A     & 20.49$^{+0.15}_{-0.15}$     & 2.46$^{+0.03}_{-0.03}$     & $-$0.95$^{+0.02}_{-0.02}$   & 0.96$^{+0.17}_{-0.15}$    & 0.32$^{+0.02}_{-0.02}$    & 0.25$^{+0.02}_{-0.02}$   & 1.05$^{+0.15}_{-0.14}$    \\
Briceño-1B     & 20.77$^{+0.13}_{-0.14}$     & 2.26$^{+0.03}_{-0.03}$      & $-$0.03$^{+0.14}_{-0.11}$    & 0.92$^{+0.14}_{-0.11}$    & 0.34$^{+0.02}_{-0.02}$    & 0.26$^{+0.02}_{-0.02}$   & 1.02$^{+0.11}_{-0.10}$     \\
Ori-East          & 28.13$^{+0.28}_{-0.23}$     & $-$1.09$^{+0.26}_{-0.22}$    & $-$1.62$^{+0.12}_{-0.12}$    & 1.48$^{+0.26}_{-0.22}$   & 1.81$^{+0.36}_{-0.24}$    & 1.11$^{+0.13}_{-0.11}$   & 2.63$^{+0.26}_{-0.25}$    \\
OBP-Far         & 30.33$^{+0.44}_{-0.45}$     & $-$2.65$^{+0.08}_{-0.08}$     & 1.94$^{+0.06}_{-0.07}$     & 3.21$^{+0.52}_{-0.42}$    & 1.03$^{+0.08}_{-0.07}$    & 0.84$^{+0.06}_{-0.06}$    & 3.48$^{+0.42}_{-0.41}$    \\
$\sigma$ Ori   & 30.94$^{+0.15}_{-0.19}$     & 2.69$^{+0.12}_{-0.14}$      & $-$0.79$^{+0.08}_{-0.08}$    & 1.30$^{+0.22}_{-0.19}$    & 1.20$^{+0.23}_{-0.16}$    & 0.97$^{+0.08}_{-0.07}$    & 2.04$^{+0.19}_{-0.18}$     \\
OBP-b             & 30.85$^{+0.62}_{-0.62}$     & $-$1.80$^{+0.04}_{-0.05}$    & $-$1.12$^{+0.04}_{-0.04}$    & 4.00$^{+0.61}_{-0.53}$    & 0.50$^{+0.03}_{-0.03}$    & 0.46$^{+0.03}_{-0.03}$   & 4.06$^{+0.57}_{-0.56}$     \\
OBP-d             & 30.79$^{+0.15}_{-0.16}$     & 0.16$^{+0.04}_{-0.03}$     & $-$0.34$^{+0.03}_{-0.03}$      & 1.05$^{+0.16}_{-0.13}$     & 0.53$^{+0.04}_{-0.04}$     & 0.54$^{+0.02}_{-0.02}$   & 1.30$^{+0.12}_{-0.11}$      \\
OBP-Near A     & 23.01$^{+0.59}_{-0.73}$     & 1.89$^{+1.53}_{-1.21}$    & $-$2.04$^{+0.03}_{-0.03}$   & 2.67$^{+1.53}_{-1.21}$    & 0.50$^{+0.12}_{-0.07}$      & 0.29$^{+0.03}_{-0.02}$    & 2.75$^{+1.28}_{-1.27}$      \\
OBP-Near B     & 21.80$^{+0.71}_{-0.60}$    & 3.02$^{+0.03}_{-0.03}$     & $-$2.15$^{+0.04}_{-0.03}$     & 2.99$^{+0.73}_{-0.56}$   & 0.29$^{+0.03}_{-0.02}$    & 0.36$^{+0.03}_{-0.03}$    & 3.02$^{+0.61}_{-0.60}$      \\
ONC                & 27.12$^{+0.09}_{-0.09}$     & 2.10$^{+0.02}_{-0.02}$     & 0.51$^{+0.03}_{-0.03}$       & 2.72$^{+0.08}_{-0.08}$      & 0.77$^{+0.02}_{-0.02}$    & 0.95$^{+0.02}_{-0.02}$     & 2.98$^{+0.07}_{-0.06}$     \\
Ori-South         & 21.42$^{+0.16}_{-0.16}$     & 0.68$^{+0.04}_{-0.04}$     & $-$1.39$^{+0.08}_{-0.08}$     & 1.60$^{+0.16}_{-0.15}$    & 0.71$^{+0.03}_{-0.03}$    & 1.17$^{+0.06}_{-0.06}$   & 2.12$^{+0.12}_{-0.11}$     \\ \hline
\end{tabular}
\end{table*}}

\section{Isochrone Fitting and Mass Distributions for OSFC Groups}
\label{app:iso_fit}
\begin{figure*}
    \centering
    \includegraphics[width=0.92\textwidth]{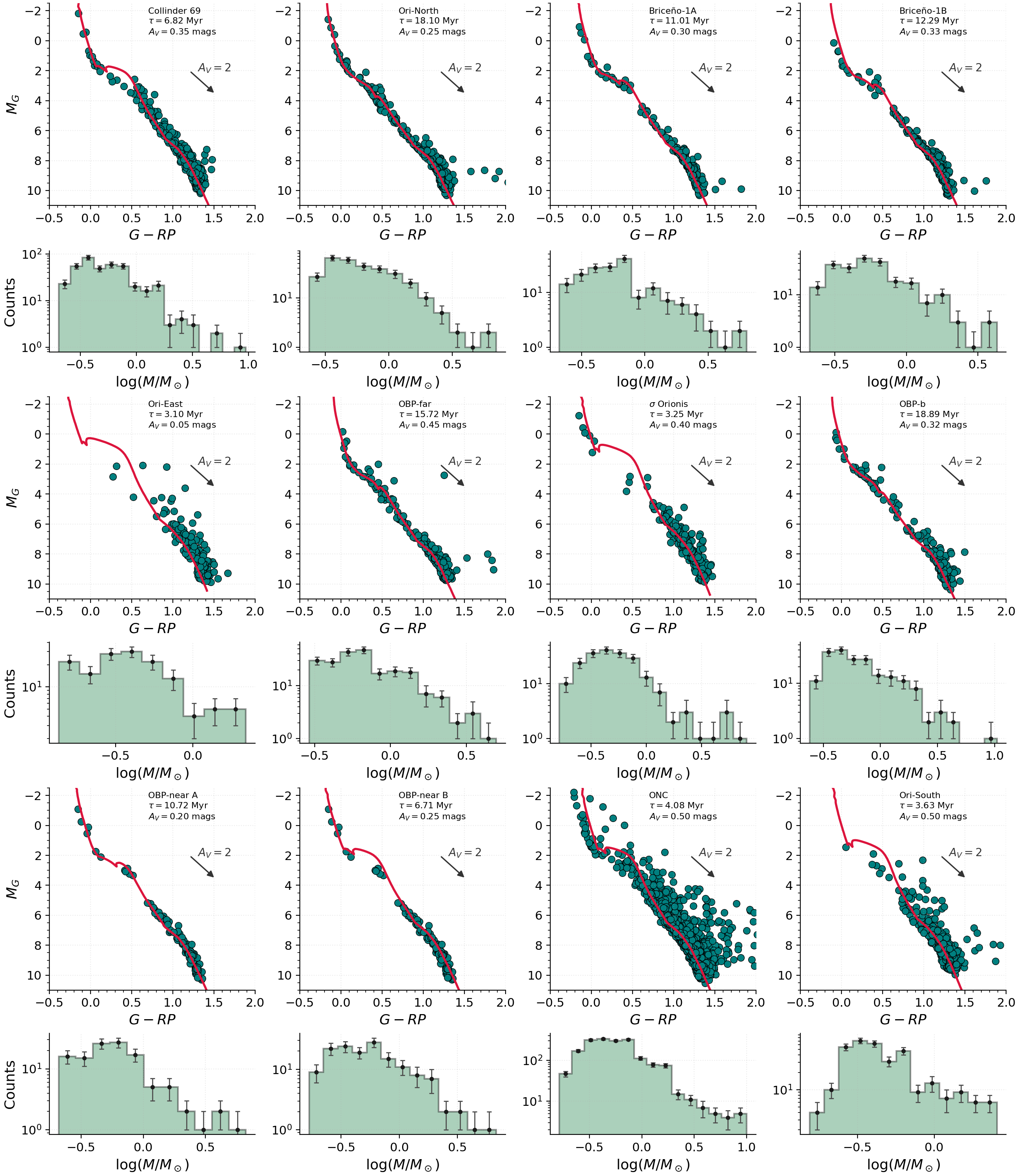}
    \caption{CMD and stellar mass distributions for 12 stellar groups in the OSFC. Each CMD shows \textit{Gaia}-DR3 photometry with the best-fitting PARSEC isochrone overlaid in red. Below each diagram, the derived mass distribution is shown with Poisson error bars. Reported values include the fitted age ($\tau$) and visual extinction ($A_V$) for each group. The ONC and Ori-East clusters display significant photometric scatter, which may bias age and extinction estimates in the fitting process.}
    \label{fig:mosaic_age_mass}
\end{figure*}

This appendix presents the CMDs and derived stellar mass distributions for the 12 additional clusters identified in the OSFC as shown in Figure~\ref{fig:mosaic_age_mass}. Each panel displays the \textit{Gaia} DR3 CMD for a given group, along with its best-fitting \texttt{PARSEC} isochrone (red line).
Below each CMD, we show the corresponding histogram of stellar masses, expressed as $\log(M/M_\odot)$, obtained through a MC sampling of colour, magnitude, age, and extinction. Poissonian errors are shown as vertical bars. Cluster age and extinction from the best fit are reported in each CMD panel.

Two cases warrant particular attention. For the ONC, the strong vertical spread in the CMD likely reflects a combination of differential extinction, variability, and binarity. As a result, the fitted isochrone may underestimate the true extinction and overestimate the cluster age. Likewise, the Ori-East group shows a sparsely populated and scattered CMD, especially at faint magnitudes. The derived age appears anomalously young given the overall distribution, suggesting either underestimated extinction or contamination by non-members.

\section{Verification of the Mass Estimation Method}
\label{app:mass_unc}

\begin{figure*}
    \centering
    \includegraphics[width=0.96\textwidth]{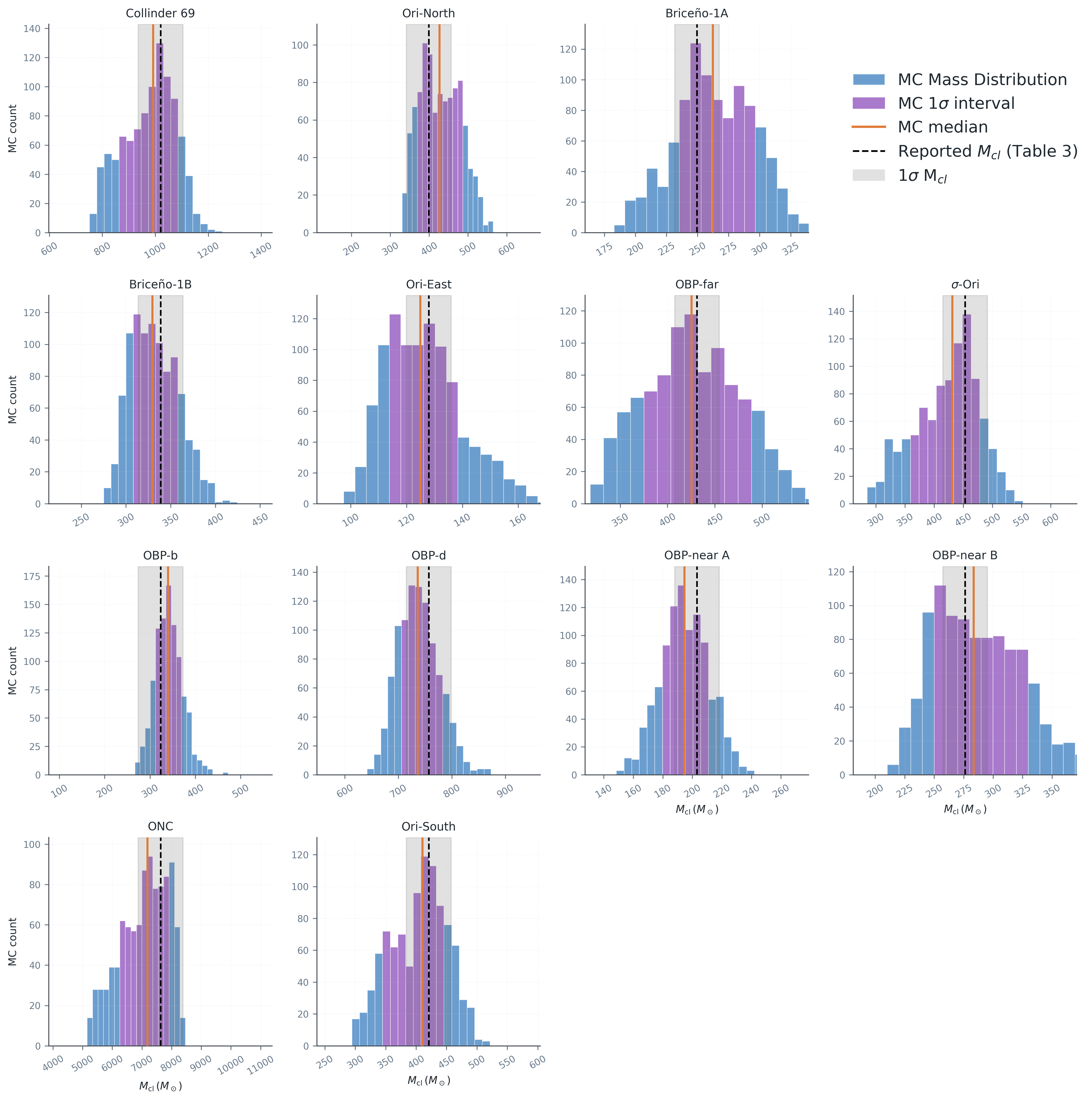}
    \caption{Distribution of the total estimated mass for each cluster from 1000 MC realizations simultaneously varying the binning configuration, and the stellar mass uncertainties. The blue histogram shows the resulting MC mass distribution, while the purple region marks the \(1\sigma\) interval around the median value. The orange vertical line indicates the median of the MC distribution, and the black dashed line corresponds to the nominal mass reported in this work. The shaded gray region represents \(1\sigma\) interval relative to the reported mass in Table~\ref{tab:params_corr}.}
    \label{fig:mc_mass}
\end{figure*}

To evaluate the sensitivity of the cluster mass estimates to the combined effects of stellar mass uncertainties and the adopted binning configuration, we performed a set of 1000 MC realizations for each cluster in our sample. In each realization, three effects were simultaneously considered: (i) the number of histogram bins, $N$, was randomly varied within $N$\(\pm1\) around the reference value determined using the Freedman--Diaconis rule, (ii) the histogram bin-edge positions were shifted uniformly within the interval \(\bmath{[-0.5,+0.5]}\) times the bin width, and (iii) the stellar masses were perturbed according to their corresponding uncertainties. Using this procedure, we recomputed the total cluster masses following the methodology adopted throughout this work.

Figure~\ref{fig:mc_mass} shows the resulting MC mass distributions. The derived median values are in good agreement with the nominal masses reported in Table~\ref{tab:params_corr},  with differences typically below 6\% and lying within the corresponding \(1\sigma\) intervals. These results indicate that the inferred cluster masses are robust to small variations in the binning configuration and stellar mass uncertainties at levels relevant for our analysis, supporting the reliability of the adopted mass-estimation methodology.

It is important to note that the cluster-mass estimation procedure is not determined by the completeness value of a single bin, but by the normalization (scaling factor) of the adopted PDMF relative to the observed mass distribution. Consequently, local fluctuations between neighbouring bins have a limited effect on the final cluster-mass estimates. This explains why the MC realizations produce only small variations in the derived cluster masses, despite local variations in the completeness-ratio trends (e.g. Figure~\ref{fig:mass_function}).

%%%%%%%%%%%%%%%%%%%%%%%%%%%%%%%%%%%%%%%%%%%%%%%%%%

% Don't change these lines
\bsp	% typesetting comment
\label{lastpage}
\end{document}